\documentclass[twocolumn,astrosymb,trackchanges]{aastex701}
\newcommand\um{\ensuremath{\mu}m}
\newcommand\kms{\ensuremath{{\rm km}~{\rm s}^{-1}}}
\newcommand\angstrom{\ensuremath{\mathring{\rm A}}}
\newcommand\ergs{\ensuremath{{\rm erg}~{\rm s}^{-1}}}

\newcommand\msun{\ensuremath{M_\odot}}
\newcommand\msunyr{\ensuremath{M_\odot~{\rm yr}^{-1}}}

\newcommand\kmsMpc{\ensuremath{{\rm km}~{\rm s}^{-1}~{\rm Mpc}^{-1}}}

\newcommand\ovi{\ion{O}{6}}
\newcommand\oviwave{\ion{O}{6}~$\lambda\lambda$1032,1038\,\angstrom}

\usepackage{threeparttable}
\usepackage{amsmath}
\usepackage{amssymb}
\usepackage{amsfonts}
\usepackage{mathbbol}
\usepackage{physics}
\usepackage{gensymb}
\usepackage{tabularx}
\usepackage{fontawesome}
\usepackage{hyperref}
\usepackage{xurl}

\let\tablenum\relax
\usepackage{siunitx}

\shorttitle{Reigniting the FUSE I}
\shortauthors{Reefe et al.}

\graphicspath{{./}{figures/}}

\begin{document}

\title{Reigniting the FUSE I: Star Formation in Massive Elliptical Galaxies}

\correspondingauthor{Michael Reefe}
\email{mreefe@mit.edu}

\author[0000-0003-4701-8497]{Michael Reefe}
\altaffiliation{National Science Foundation, Graduate Research Fellow}
\affiliation{Kavli Institute for Astrophysics \& Space Research, Massachusetts Institute of Technology, Cambridge, MA 02139, USA}
\email{mreefe@mit.edu}

\author[0000-0001-5226-8349]{Michael McDonald}
\affiliation{Kavli Institute for Astrophysics \& Space Research, Massachusetts Institute of Technology, Cambridge, MA 02139, USA}
\email{mcdonald@space.mit.edu}

\author[0000-0002-3031-2326]{Eric Miller}
\affiliation{Kavli Institute for Astrophysics \& Space Research, Massachusetts Institute of Technology, Cambridge, MA 02139, USA}
\email{milleric@space.mit.edu}

\author[orcid=0000-0002-1793-9968]{Thomas Brown}
\affiliation{Space Telescope Science Institute, Baltimore, MD 21218, USA}
\email{tbrown@stsci.edu}

\begin{abstract}

In this inaugural paper of our ``Reigniting the \textit{FUSE}'' series, we present the data collection \& reduction, methodology, and preliminary results for a comprehensive and homogeneous reanalysis of all 30 massive elliptical galaxies that have archival far-ultraviolet (FUV) spectra with the \textit{Far Ultraviolet Spectroscopic Explorer} (\textit{FUSE}).  We perform SED fitting to these spectra, with aperture-matched photometry from \textit{GALEX}, SDSS, DES/DECaLS, 2MASS, and \textit{WISE}, and construct a catalogue of high quality constraints on their stellar masses, star formation histories, stellar and gas-phase metallicities, and dust opacities.  As a part of this analysis, we have also compiled a comprehensive modern set of stellar template libraries, with high resolution in the FUV, and broad coverage from $90\,\angstrom$--$1000\,$\um~ at lower resolution, which is made available through modified versions of \textsc{FSPS} and \textsc{Python FSPS}.  We find strong evidence to suggest that the youngest stars in these systems do not form out of recycled gas from the old stellar populations, but instead are fueled either by gas cooling out of the hot circumgalactic/intracluster medium, or cool gas stripped from nearby or merging dwarf systems.  Future papers in this series will reexamine the \oviwave\ emission in the context of the cooling flow problem.

\end{abstract}

\section{Introduction} \label{sec:intro}

All massive galaxies host a halo of hot, diffuse gas in their circumgalactic medium (CGM) with characteristic temperatures of $T \sim 10^7$ K and densities of $n \sim 0.01$ cm$^{-3}$. In some of the most massive elliptical galaxies in the universe, sitting at the centers of deep potential wells in galaxy clusters and groups, their CGM is proportionally massive, and it may become dense enough that it should cool down within less than a Hubble time, $t_{\rm cool} \propto T^{1/2}n^{-1}$ \citep[see review by][]{1994ARA&A..32..277F}.  In these ``cool core'' systems, this new reservoir of cold gas can then act as fuel for new star formation, breathing new life into what would otherwise be classified as ``red and dead'' galaxies.

The rate that the hot CGM cools and flows inwards ($\dot{M}_c$) can be related to the amount of energy it loses through X-ray radiation per unit time ($L_X$), and the thermal energy/enthalpy per unit mass ($\sim kT_X/\mu m_p$), such that
\begin{equation}
    \dot{M}_{\rm c} = \alpha\frac{\mu m_p}{ kT_X}L_X~,
\end{equation}
where the numerical proportionality $\alpha$ depends on if the cooling is isobaric (at constant pressure, $\alpha = 2/5$) or isochoric (at constant density, $\alpha = 2/3$).  Importantly, this estimate, often called the ``classical'' cooling rate, assumes that the gas is cooling in a steady-state at the current $L_X$ over at least a cooling time, and that the gas is not reheated \citep{1988xrec.book.....S}.  Therefore, for practical purposes, it often acts as a maximum cooling rate when compared to other estimates.

Nevertheless, important information can be learned about the state of a system by comparing the classical cooling rate to the star formation rate (SFR).  Under the assumption that the star formation is fueled by the cooling gas, the ``cooling efficiency'' $\eta \equiv {\rm SFR}/\dot{M}_c$ measures the fraction of cooling gas that turns into stars.  In most systems, $\eta \sim 1$--$10\%$ \citep{1985ApJS...59..447H, 1987MNRAS.224...75J, 1989ApJ...338...48H, 1989AJ.....98.2018M, 1999MNRAS.306..857C, 2008ApJ...681.1035O, 2010ApJ...721.1262M, 2012ApJS..199...23H, 2018ApJ...858...45M, 2023ApJ...947...44C}, indicating that a large majority of the gas gets reheated or otherwise is prevented from collapsing. This inefficiency is thought to be driven primarily by feedback from an active galactic nucleus (AGN), observed in the form of bipolar radio jets, which inject turbulence and inflate large buoyant bubbles of plasma in the surrounding CGM \citep{2000MNRAS.318L..65F, 2001ApJ...554..261C, 2004ApJ...607..800B, 2005Natur.433...45M, 2005MNRAS.364.1343D, 2006ApJ...652..216R, 2007ApJ...665.1057F, 2008ApJ...686..859B, 2012MNRAS.421.1360H, 2015ApJ...805...35H, 2023ApJ...954...56O}.  The mechanical power within these radio jets is enough to offset the total $L_X$ in all but the brightest quasar systems \citep{2013MNRAS.432..530R}.  The residual cooling seen at the 1\% level may develop from local thermodynamic instabilities, despite this global balance, the behavior of which can be reproduced in hydrodynamic simulations in the right conditions \citep[e.g.][]{2012ApJ...746...94G, 2013MNRAS.432.3401G, 2012MNRAS.419.3319M}.  For reviews, see \citet{2007ARA&A..45..117M, 2012ARA&A..50..455F, 2012NJPh...14e5023M, 2020NatAs...4...10G, 2022PhR...973....1D}.

This picture, while compelling, leaves many details of the balance between feedback and cooling uncertain.  Where and how local thermal instabilities develop, and how the CGM is heated isotropically, are crucial areas of uncertainty \citep[see, e.g.,][]{2016ApJ...829...90Y}, as many cooling rate and SFR measurements lack spatial information.  In the time domain, the duty cycle of the AGN jets is only loosely constrained.  \citet{2012MNRAS.427.3468B} used the incidence rate of bubbles in a sample of massive galaxies to put a lower limit on the population-wide duty cycle of $\gtrsim 60\%$. Studies can also be performed on individual systems which show evidence for multiple cavities in the hot CGM, relating to multiple past episodes of feedback \citep[e.g.][]{2020A&A...638A..29B, 2021A&A...650A.170B, 2021ApJ...923L..25U, 2025A&A...696A.239B}.  These have found evidence for duty cycles approaching 100\%, consistent with \citet{2012MNRAS.427.3468B}, indicating almost continuous jet activity.  While these are important clues, individual systems where this measurement is possible are few and far between.  Finally, it is often taken for granted that the observed SFR is being fueled by cooling gas from the CGM, but cool gas may also be accreted from nearby or merging systems \citep[e.g.][]{2022A&A...666A..94O} or recycled from old stellar populations undergoing mass loss \citep{2011ApJ...738L..24V, 2018ApJ...858...45M}.

With these caveats in mind, it becomes necessary to obtain more accurate measurements of the cooling rate at lower temperatures to obtain a physical picture of how cooling flows proceed.  This can be done using spectroscopic line transitions, which are sensitive to a narrow temperature range of the cooling function, and reflect cooling on the recombination timescale $t_{\rm rec}$, which is generally much shorter than $t_{\rm cool}$.  For example, X-ray spectroscopy has been used to measuring cooling through $\sim$$10^7$ K with lines like \ion{O}{7}--\ion{O}{8}, \ion{Ne}{9}--\ion{Ne}{10}, \ion{Mg}{11}--\ion{Mg}{12}, \ion{Fe}{17}--\ion{Fe}{24}, among others \citep{2001A&A...365L..99K, 2001A&A...365L..87T, 2001ApJ...560..194M, 2001A&A...365L.104P, 2003ApJ...590..207P, 2018MNRAS.480.4113P}.  See review by \citet{2006PhR...427....1P}.  UV/optical/IR spectroscopy can probe $\sim$$10^{5.5}$ K with ``coronal'' lines like \ion{O}{6} \citep{2001ApJ...560..187O, 2004A&A...421..503L, 2005ApJ...635.1031B, 2006ApJ...642..746B, 2014ApJ...791L..30M, 2015ApJ...811..111M, 2017ApJ...835..216D}, [\ion{Fe}{10}] \citep{2011MNRAS.417.3080C}, and [\ion{Ne}{6}] \citep{2025Natur.638..360R}.

X-ray spectroscopy measurements, however, have produced perplexing results, with X-ray cooling rates $\dot{M}_X$ often lower than SFRs, even when accounting for large systematic uncertainties \citep{2016A&A...595A.123M}. There are three possibilities to explain this, either: (1) most of the gas does not cool below $10^7$ K, and the observed SFR is not fueled by cooling gas from the CGM; (2) the observed $\dot{M}_X$ is lower than the true $\dot{M}_X$ due to absorption from neutral hydrogen along the line of sight; or (3) the SFRs are systematically overestimated.  Option (2) has become known as the ``hidden cooling flow'' theory \citep{2022MNRAS.515.3336F, 2023MNRAS.521.1794F, 2023MNRAS.524..716F}, and may be further explored with next generation X-ray microcalorimetry.  Option (3) may be diagnosed with more precise star formation rate metrics, such as with far-UV spectroscopy. 

Measurements from UV/optical coronal lines do not observe the same perplexingly low signal, and are often somewhere in between the SFR and $\dot{M}_c$, which lends credibility to the ``hidden cooling flow'' theory.  In particular, far-UV spectroscopy has been obtained for 30 massive elliptical systems with the \textit{Far Ultraviolet Spectroscopic Explorer} (\textit{FUSE}) mission \citep{2000ApJ...538L...1M}, with \oviwave\, detections in 14 \citep{2001ApJ...560..187O, 2004A&A...421..503L, 2005ApJ...635.1031B, 2006ApJ...642..746B}. These studies, while important in definitively establishing the presence of residual cooling flows, have been performed inhomogeneously and with now-outdated analysis techniques that make them difficult to interpret and compare to more modern measurements.  In particular:
\begin{itemize}
\item The majority of published \ovi\, luminosities are based on distances which are now known to be incorrect by factors of $\sim 15$--$40$\%.  \citet{2001ApJ...560..187O} and \citet{2005ApJ...635.1031B} assume a Hubble constant of 50 \kmsMpc, having implications for the more distant systems, while the more nearby ellipticals studied in \citet{2005ApJ...635.1031B} use redshift-independent distances which have largely been superseded by more accurate measurements.
\item While all published \ovi\, luminosities account for Galactic extinction, only \citet{2001ApJ...560..187O} considers the effects of intrinsic extinction.  Given the typical levels of extinction in massive elliptical galaxies \citep[$E(B-V) \sim 0.1$ mag;][]{1999MNRAS.306..857C}, this would lead to typical attenuation correction factors of $\sim 3$--$4$.
\item UV continuum modeling in these studies was done either with a simple linear function \citep{2001ApJ...560..187O, 2004A&A...421..503L} or by using NGC 1399 (Fornax A) as a template \citep{2005ApJ...635.1031B, 2006ApJ...642..746B}.  Differences in assumptions about the underlying continuum can lead to large changes in the inferred line flux, of up to 50\% \citep[as noted by][]{2005ApJ...635.1031B}.
\item The conversion of an \ovi\, luminosity to a cooling rate was done using a standard conversion of $(\dot{M}_{\rm O\,VI}/\msunyr) = 1.11 \times 10^{-39} (L_{\rm O\,VI}/\ergs)$ from \citet{1986ApJ...310L..27E}, which does not account for the unique geometry, thermodynamics, and metal enrichment of the gas in each system.
\end{itemize}

In this series of papers, we perform a detailed multiwavelength reassessment of the cooling flow problem across the cluster, group, and elliptical scales, with reanalyzed archival \textit{FUSE} data of these 30 massive elliptical galaxies as a cornerpiece.  This is the first time that all of these systems have been considered together in a homogeneous analysis.  The reanalyzed \textit{FUSE} spectra not only provide an updated look into \ovi\, cooling rates, but being in the far-UV regime, also provide strong constraints on the ages and metallicities of the youngest stars with stellar population synthesis models.  The far-UV suffers from intense dust extinction, but this effect can be mitigated by including aperture-matched broadband photometry spanning the UV, optical, and IR, which allows us to include dust reddening in our models.  These results can then be compared to aperture-matched X-ray analyses with the \textit{Chandra X-ray Observatory} to trace the cooling flow across multiple gas phases.

Paper I (this paper) lays out our methodology for the \textit{FUSE} data reduction, calculating aperture-matched photometry and \textit{Chandra} spectra, and SED continuum modeling. We present our first results on the stellar populations, star formation histories, metallicities, and dust contents.  With the continuum modeling in place, paper II will then cover our analysis of the \ovi\, emission, comparing the \ovi-derived cooling rates to a homogeneous recalculated set of classical cooling rates from \textit{Chandra} data, and in turn comparing them to the SFRs from paper I.  In future publications, we may even further build on these results by modeling the \ovi\, emission simultaneously with data from the \textit{X-ray Multi-Mirror Mission}'s Reflection Grating Spectrograph (\textit{XMM}-RGS) covering \ion{O}{7} and \ion{O}{8}, allowing for abundance-independent and absorption-corrected cooling measurements over 2 orders of magnitude in temperature.

The organization of this paper is as follows: $\S$\ref{sec:reduce} covers the \textit{FUSE} data reduction and collection of aperture-matched broadband photometry and \textit{Chandra} spectra; $\S$\ref{sec:analysis} details our techniques for modeling the UV--IR spectral energy distribution (SED) continuum with stellar population synthesis, and modeling the \textit{Chandra} spectra to obtain CGM gas metallicities; $\S$\ref{sec:results} presents our results on the stellar population properties, star formation histories, and dust contents; $\S$\ref{sec:discussion} discusses the implications of these results in the context of the cooling flow problem; and $\S$\ref{sec:conclusion} gives our closing remarks and takeaways.  Throughout this paper, we assume a flat $\Lambda$CDM cosmology with $H_0=70$ km s$^{-1}$ Mpc$^{-1}$, $\Omega_m = 0.27$, and $\Omega_\Lambda = 0.73$.  Uncertainties are $1\sigma$ (68\% confidence) unless otherwise stated.

\section{Data Reduction} \label{sec:reduce}

Details of the 29 massive elliptical galaxies in our sample, including coordinates, redshifts, and distances, are given in Table \ref{tab:info}.  One of the 30 total aforementioned systems observed with \textit{FUSE}, NGC 3585 had to be removed from our analysis due to updates to the bad data flagging in the CalFUSE reduction pipeline rendering none of its 16 ks of data usable.  Therefore, our final sample consists of the remaining 29 systems.

\begin{deluxetable*}{llDDDCLll}
\def\arraystretch{0.85}
\tabletypesize{\footnotesize}
\tablecaption{Target observational information for our sample.}
\label{tab:info}
\tablehead{
    \colhead{Target Name} & \colhead{Alternate Name} & \twocolhead{R.A. (deg.)} & \twocolhead{Dec. (deg.)} & \twocolhead{Redshift} & \colhead{$D$ (Mpc)\tablenotemark{a}} & \colhead{$N_{\rm p}$\tablenotemark{b}} & \colhead{CXO Bg.\tablenotemark{c}} & \colhead{Optical Survey\tablenotemark{d}}
}
\decimals
\startdata
\multicolumn{12}{c}{Cluster Halos}\\
\hline
Perseus (Abell 426)  & NGC 1275       & 49.95067    & 41.51170   & 0.017670  & 69.0          \pm 3.2 & 1 & BL & SDSS   \\
Fornax (Abell S0373) & NGC 1399       & 54.62092    & -35.45019  & 0.004753  & 17.7          \pm 0.4 & 1 & BL & DES    \\
Virgo                & M 87           & 187.70593   & 12.39112   & 0.004283  & 16.8          \pm 0.3 & 2 & BL & SDSS   \\
Abell 1795           & CGCG 162-010   & 207.21872   & 26.59289   & 0.063309  & 284.5^{\bf *}         & 2 & BL & SDSS   \\
Abell 2029           & IC 1101        & 227.73375   & 5.74479    & 0.078054  & 354.6^{\bf *}         & 1 & BL & SDSS   \\
Abell 2597           & PKS 2322-12    & 351.33215   & -12.12417  & 0.082991  & 378.3^{\bf *}         & 1 & BL & DECaLS \\
Abell 3112           & ESO 248- G 006 & 49.49035    & -44.23811  & 0.076106  & 345.2^{\bf *}         & 1 & BL & DES    \\
AWM 7                & NGC 1129       & 43.61413    & 41.57954   & 0.017472  & 75.9^{\bf *}          & 1 & BL & SDSS   \\
\hline
\multicolumn{12}{c}{Group Halos}\\
\hline
IC 1459  &       & 344.29420 & -36.46222 & 0.006011 & 26.2 \pm 2.1  & 1 & OS & DECaLS \\
NGC 1316 &       & 50.67375  & -37.20794 & 0.006010 & 19.2 \pm 0.6  & 1 & OS & DES    \\ 
NGC 1395 &       & 54.62447  & -23.02747 & 0.005727 & 22.8 \pm 1.1  & 1 & OS & DES    \\
NGC 1407 &       & 55.04934  & -18.58004 & 0.005934 & 24.1 \pm 1.5  & 1 & OS & DES    \\
NGC 3115 &       & 151.30824 & -7.71858  & 0.002287 & 10.2 \pm 0.5  & 1 & OS & DECaLS \\
NGC 3379 & M 105 & 161.95669 & 12.58160  & 0.003026 & 11.1 \pm 0.3  & 1 & OS & DECaLS \\
NGC 3607 &       & 169.22778 & 18.05178  & 0.003102 & 21.4 \pm 1.4  & 1 & OS & SDSS   \\
NGC 3923 &       & 177.75713 & -28.80594 & 0.005801 & 21.6 \pm 1.6  & 1 & OS & DECaLS \\
NGC 4125 &       & 182.02507 & 65.17413  & 0.004273 & 22.8 \pm 1.8  & 1 & OS & SDSS   \\
NGC 4636 &       & 190.70761 & 2.68780   & 0.003129 & 16.3 \pm 0.6  & 2 & BL & SDSS   \\
NGC 4649 & M 60  & 190.91665 & 11.55261  & 0.003703 & 16.7 \pm 0.4  & 2 & BL & SDSS   \\
NGC 5846 &       & 226.62180 & 1.60629   & 0.005711 & 28.8 \pm 2.0  & 1 & BL & SDSS   \\
\hline
\multicolumn{12}{c}{Elliptical Halos}\\
\hline
NGC 1404 &      & 54.71638  & -35.59411 & 0.006448  & 19.3 \pm 0.4 & 1 & BL & DES    \\
NGC 1549 &      & 63.93794  & -55.59218 & 0.004190  & 16.6 \pm 1.0 & 1 & OS & DES    \\
% NGC 3585 &      & 168.32127 & -26.75485 & 0.004783  & 17.1 \pm 1.1 & 1 &  & DES    \\
NGC 4374 & M 84 & 186.26560 & 12.88698  & 0.003392  & 17.0 \pm 0.4 & 1 & OS & SDSS   \\
NGC 4406 & M 86 & 186.54896 & 12.94619  & -0.000747 & 16.4 \pm 0.4 & 1 & BL & SDSS   \\
NGC 4472 & M 49 & 187.44484 & 8.00048   & 0.003272  & 16.1 \pm 0.3 & 1 & BL & SDSS   \\
NGC 4494 &      & 187.85043 & 25.77525  & 0.004476  & 13.4 \pm 0.5 & 1 & OS & SDSS   \\
NGC 4552 & M 89 & 188.91586 & 12.55634  & 0.001134  & 16.5 \pm 0.5 & 1 & OS & SDSS   \\
NGC 4621 & M 59 & 190.50940 & 11.64695  & 0.001558  & 16.0 \pm 0.4 & 1 & OS & SDSS   \\
NGC 5102 &      & 200.49003 & -36.63024 & 0.001556  & 10.2 \pm 3.6 & 1 & OS & DECaLS \\
\hline
\enddata
\tablenotetext{}{Note: All coordinates, redshifts, and distances have been retrieved from the NASA/IPAC Extragalactic Databse (NED): \href{https://ned.ipac.caltech.edu/}{https://ned.ipac.caltech.edu/}.}
\tablenotetext{a}{Mean redshift-independent distance}
\tablenotetext{b}{Number of unique \textit{FUSE} pointings}
\tablenotetext{c}{The type of background used to extract \textit{Chandra} spectra. ``BL'' for blank sky backgrounds, and ``OS'' for off-source backgrounds from the same data.}
\tablenotetext{d}{The optical all-sky survey that was used to obtain optical photometry}
\tablenotetext{*}{These systems use the redshift-derived luminosity distance, using our standard cosmology.}
\end{deluxetable*}

\subsection{\textit{FUSE} Spectroscopy} \label{sec:fuse}

We retrieve the raw \textit{FUSE} data products for our sample from the Mikulski Archive for Space Telescopes (MAST)\footnote{\url{https://mast.stsci.edu/portal/Mashup/Clients/Mast/Portal.html}}.  The data are observed using the low-resolution (LWRS) $30'' \times 30''$ aperture. Most of the galaxies in our sample have a single (central) pointing, while Virgo, Abell 1795, NGC 4636, and NGC 4649 each have two pointings.  In these cases, each pointing is reduced and analyzed separately.  The data are reduced using the most up-to-date version of the CalFUSE pipeline \citep[v3.2.3;][]{2007PASP..119..527D}.

Previous studies analyzing these data have used a restricted pulse height range \citep[i.e. 4--15;][]{2001ApJ...560..187O} to reduce backgrounds.  However, more recent guidelines from the CalFUSE developers suggest using more inclusive limits ($\sim$2--24) to avoid substantial flux loss, which can present itself as apparent absorption features in the output spectrum \citep{2007PASP..119..527D}.  We therefore use these more inclusive limits for our analysis.  In addition, we use only data observed during orbital night in order to reduce contamination from terrestrial airglow lines.

Once each individual exposure is reduced, we use the FUSE Tools in C\footnote{\url{https://archive.stsci.edu/fuse/analysis/toolbox.html}} to combine them into a single spectrum (per detector, per pointing) following the recommendations for faint targets listed in the online FUSE handbook\footnote{\url{https://archive.stsci.edu/fuse/DH_Final/}}.  This consists of combining the intermediate data files with \texttt{idf\_combine}, combining bad pixel maps with \texttt{bpm\_combine}, and then extracting spectra with \texttt{cf\_extract\_spectra}.  This optimizes the estimation of the background emission. Despite following this procedure, we find that due to the especially faint nature of many of our targets, the background is often overestimated.  As such, we disable the subtraction of the background spectrum and instead save both the total output spectrum and the background spectrum separately, so that we may include the background as a component during our continuum fitting anlaysis ($\S$\ref{sec:continuum}).

For each pointing, we then combine the spectra from each of \textit{FUSE}'s 8 detectors into a single composite spectrum by taking the spectrum with the lowest noise level at each wavelength sample.  This almost always results in the LiF 1a spectrum being taken for the wavelength range that the \ovi~ doublet falls within.  

Given the faint extended targets and higher-than-necessary spectral resolution of \textit{FUSE} ($R \sim 20,000$), the output spectra are highly Poissonian.  We therefore bin the final spectra to a wavelength resolution of $\sim 0.2~\angstrom$, or $\sim 60~\kms$ at $1000~\angstrom$.  This ensures that every wavelength sample has data, while preserving a fine enough velocity resolution to measure typical velocities seen in the CGM.

\subsection{Photometry} \label{sec:photometry}

We obtain sky cutouts and calculate aperture-matched photometry (matched to the \textit{FUSE} LWRS aperture) in up to 14 bands per target. These bands cover the UV with the \textit{Galaxy Evolution Explorer} \citep[\textit{GALEX};][]{2005ApJ...619L...1M}; the optical with the Sloan Digital Sky Survey \citep[SDSS;][]{1998AJ....116.3040G,2006AJ....131.2332G,2017AJ....154...28B,2022ApJS..259...35A}, the Dark Energy Survey \citep[DES;][]{2016MNRAS.460.1270D,2021ApJS..255...20A}, and the Dark Energy Camera Legacy Survey \citep[DECaLS;][]{2019AJ....157..168D}; the near-IR with the Two Micron All Sky Survey \citep[2MASS;][]{2006AJ....131.1163S}; and the mid-IR with the \textit{Wide-field Infrared Survey Explorer} \citep[\textit{WISE};][]{2010AJ....140.1868W,2011ApJ...731...53M}.

\subsubsection{GALEX} \label{sec:galex}

\textit{GALEX} FUV and NUV band sky cutouts are extracted from MAST, and individual exposures are combined into a composite image using the exposure times as weights.  Count rates are converted into physical flux units using the calibrations reported in \citet{2007ApJS..173..682M} and on the GALEX Guest Investigator website\footnote{\url{https://asd.gsfc.nasa.gov/archive/galex/FAQ/counts\_background.html}}.  Statistical uncertainties, on a per-pixel level, are estimated by taking the square root of the raw counts per second in the non-background-subtracted images.  For systematic uncertainties, a 10\% absolute flux calibration uncertainty is assumed, along with a repeatability uncertainty of 0.050 mag in the FUV band and 0.027 mag in the NUV band \citep{2007ApJS..173..682M}. 

\subsubsection{SDSS} \label{sec:sdss}

For northern targets, SDSS DR17 u, g, r, i, and z band sky cutouts are retrieved from the SDSS Science Archive Server\footnote{\url{https://data.sdss.org/sas/}}.  The images are already combined and flux calibrated, so no additional data processing is necessary.  To calculate statistical errors, we follow the instructions on the SDSS Data Model archive\footnote{\url{https://data.sdss.org/datamodel/files/BOSS\_PHOTOOBJ/frames/RERUN/RUN/CAMCOL/frame.html}} to calculate the raw data counts, including sky and dark current, and take the square root.  We assume a systematic calibration error of 0.04 mag in the u band and 0.02 mag in the z band based on measured offsets from the AB system as noted on the SDSS website\footnote{\url{https://www.sdss4.org/dr16/algorithms/fluxcal}} \citep[see also][]{2006ApJS..167...40E,2007AJ....134..973I}, and a conservative repeatability error of 2\% in all bands \citep{2008ApJ...674.1217P}.

\subsubsection{DES \& DECaLS} \label{sec:des}

For southern targets, DES DR2 g, r, i, z, and Y band sky cutouts are obtained from the NOIRLab Astro Data Archive's\footnote{\url{https://astroarchive.noirlab.edu/}} Simple Image Access service, through the ``coadd\_all'' endpoint.  The images are calibrated to absolute flux units using the AB magnitude zero points stored in the image headers.  Statistical errors are estimated using the raw image counts, with an estimate of the median sky counts added from \citet{2021ApJS..255...20A}.  For systematic uncertainties, we assume a conservative calibration uncertainty of 0.02 mag \citep{2021ApJS..255...20A,2018AJ....155...41B}.  In some of our optically bright targets, the DES images become saturated.  If we detect more than 2\% of the pixels in the aperture extraction region to be saturated, we consider the photometry unreliable and mask it out.

If no image is available from the DES, we also search for images from DECaLS in the g, r, i, and z bands through the NERSC Cosmology Data Repository\footnote{\url{https://portal.nersc.gov/cfs/cosmo/}}.  DECaLS cutouts are already flux calibrated, and statistical uncertainties are provided, so no additional data processing is necessary, though they exhibit the same saturation issues as the DES image.  We assume the same systematic uncertainties as we do for the DES images.

\subsubsection{2MASS} \label{sec:2mass}

2MASS J, H, and K$_{\rm s}$ band sky cutouts are downloaded from the NASA/IPAC Infrared Science Archive (IRSA)\footnote{\url{https://irsa.ipac.caltech.edu/frontpage/}}.  Sky subtraction is done first using the sky value in the image header.  The counts are then converted into magnitudes on the 2MASS system using the magnitude zero point, also stored in the header. We use the absolute flux calibration of \citet{2003AJ....126.1090C} to convert these magnitudes into physical flux values.  Statistical uncertainties are estimated using the formula given by the online 2MASS documentation\footnote{\url{https://irsa.ipac.caltech.edu/data/2MASS/docs/releases/allsky/doc/sec6\_8a.html}}, which accounts for the standard Poissonian noise (including read noise), with additional corrections for correlated noise in resampled pixels, smoothing from the image co-addition kernel, and background fitting for extended sources.  We assume a $\sim$10\% systematic uncertainty in the absolute flux calibration and a uniformity error of 0.044 mag\footnote{\url{https://www.ipac.caltech.edu/2mass/releases/allsky/doc/sec6\_1.html}}.  

\subsubsection{WISE} \label{sec:wise}

\textit{WISE} 3.4 \um~(W1), 4.6 \um~(W2), 12 \um~(W3), and 24 \um~(W4) band sky cutouts are downloaded from the AllWISE Data Release on IRSA.  We estimate a 2D background for these images using a 3$\sigma$-clipped median within boxes of $500 \times 500$ pixels, gridded over the full image, which are then smoothly interpolated to create a background image.  This is done using the \texttt{Background2D} routine in python's \textsc{Photutils} package.  The background-subtracted images are then calibrated using the absolute flux calibrations calculated by \citet{2011ApJ...735..112J}. Statistical errors are calculated following the recommended procedure on the online AllWISE Explanatory Supplement\footnote{\url{https://wise2.ipac.caltech.edu/docs/release/allwise/expsup/sec4\_3a.html}}, which includes standard Poissonian noise, as well as uncertainty in the background estimation procedure.  Systematic uncertainties are calculated based on errors in the absolute flux calibration, which are on the level of 1--2\%, in addition to a $\sim$3\% repeatability error (all of which are listed in the Explanatory Supplement).

For the longer-wavelength surveys (2MASS and \textit{WISE}), confusion noise can become an increasingly important source of uncertainty on absolute flux measurements due to the increasing PSF size and decreasing pixel resolution.  As such, we also include an estimate of the confusion noise in our photometry measurements for these surveys.  The confusion noise is calculated as\footnote{\url{https://wise2.ipac.caltech.edu/docs/release/allwise/expsup/sec4\_3a.html}}:
\begin{equation}
    \sigma_{\rm conf} = \sqrt{\sigma_{\rm apbk}^2 - f_{\rm corr}N_{\rm A}\langle \sigma_{i}^2\rangle_{\rm bk}}
\end{equation}
where $\sigma_{\rm apbk}^2$ is the variance in the background level on the scale of the measurement aperture, $\langle \sigma_{i}^2\rangle_{\rm bk}$ is the median per-pixel variance in the background aperture, $N_{\rm A}$ is the number of pixels in the measurement aperture, and $f_{\rm corr}$ is a correlated noise correction factor.  To estimate $\sigma_{\rm apbk}^2$, we place a series of apertures (of the same size and shape as our target aperture) evenly over the whole image, spaced out by 5 pixels.  We throw out any apertures that are contaminated by point sources, which we determine using \textsc{Photutils}'s \texttt{find\_peaks} algorithm with a $3\sigma$ detection threshold.  We then measure the variance in the background level measured from each of the remaining apertures.  Values for $f_{\rm corr}$ are provided for \textit{WISE}, while for 2MASS we assume $f_{\rm corr} = 4$ based on the ratio of pixel scales in the single-frame and coadded images.

\subsection{\textit{Chandra} Imaging Spectrsocopy} \label{sec:chandra}

We use the \textit{Chandra} Interactive Analysis of Observations (CIAO) software to retreive and process \textit{Chandra} data \citep{2006SPIE.6270E..1VF}.  For each object in our sample, we use \texttt{find\_chandra\_obsid} to search for all Advanced CCD Imaging Spectrometer (ACIS) observations taken with no grating and which are cospatial with the \textit{FUSE} pointing's coordinates within a 30$''$ radius.  We retrieve all such ObsIDs and process them with \texttt{chandra\_repro}. 

For each ObsID, we extract a spectrum and all relevant response files from the \textit{FUSE} LWRS aperture using the \texttt{specextract} function.  Due to the large variation in exposure time and in the targets' intrinsic brightness, some observations are suitable to use an off-source aperture to extract a background spectrum, and others are not.  For targets with small angular sizes where this is possible, we extract a background spectrum and response files from a source-free aperture on the same chip.  For the larger targets, we instead use matched ``blank sky'' backgrounds, retrieved using the \texttt{blanksky} function on data which has been deflared.  We record which type of background is used for each source in Table \ref{tab:info}.

The deflaring process is done in a few steps.  First, we use \texttt{vtpdetect} on the reprocessed event file, after binning by a factor of 8 in x and y, to detect point and extended sources in the image.  These are then masked out, and a light curve of the background, binned to 259.28s, is obtained with \texttt{dmextract}.  We then use \texttt{deflare} on the light curve with \texttt{method=clean}.  The resultant GTI file is then used as a mask in the time dimension to exclude time intervals where the background is flaring.  This masked event file is finally used as the input for the \texttt{blanksky} function to obtain the best possible match for blank sky backgrounds.  Corresponding response files are not obtained for blank sky backgrounds, since they are aggregates of many observations (and the backgrounds are extremely sub-dominant in these sources, especially within the central \textit{FUSE} apertures that we extract spectra from).

Point sources, including nuclear AGN when relevant, are manually masked out from the spectral extraction regions and background regions in the images. Finally, source spectra, background spectra, and response files from each ObsID are combined with \texttt{combine\_spectra} to create a single composite spectrum for each source.

\section{Spectral Fitting Methods} \label{sec:analysis}

Using the \textit{FUSE} spectra in combination with the aperture-matched broadband photometry, we perform joint spectrophotometric SED fitting with the \textsc{Prospector} python library \citep{2021ApJS..254...22J}, which is built on the Flexible Stellar Population Synthesis \citep[FSPS;][]{2009ApJ...699..486C,2010ApJ...712..833C} framework for stellar continuum modeling.  FSPS, and by extension \textsc{Prospector}, are highly flexible in allowing different prescriptions for star formation histories, initial mass functions, dust attenuation and reprocessed emission, and calibration components when jointly fitting spectroscopy and photometry.  

\subsection{Stellar Template Libraries} \label{sec:templates}

\begin{deluxetable}{lDccl}
\def\arraystretch{0.85}
\tabletypesize{\footnotesize}
\tablecaption{Stellar template library wavelength and metallicity coverage.}
\label{tab:templatelibs}
\tablehead{
    \colhead{Code} & \twocolhead{$\Delta\lambda$\tablenotemark{a}} & \colhead{$\lambda_{\rm min/max}$\tablenotemark{b}} & \colhead{$Z_{\rm min/max}$\tablenotemark{c}} & \colhead{Ref.} \\
    \colhead{} & \twocolhead{($\angstrom$)} & \colhead{($\angstrom$)} & \colhead{($\log(Z/Z_\odot)$)} & \colhead{}
}
\decimals
\startdata
\hline
\textsc{Phoenix}    & \lesssim 0.1  & $(0, 10^7)$     & $(-4.0, +0.5)$ & A12, A14 \\
\textsc{Tlusty}     & 0.1           & $(90, 10^6)$    & $(-1.9, +0.1)$ & B96 \\
\textsc{PoWR} (OB)  & 0.01          & $(5, 10^9)$     & $(-1.5, +0.0)$ & H19 \\
\textsc{PoWR} (WR)  & 0.1           & $(5, 10^9)$     & $(-1.2, +0.0)$ & T15, S12 \\
\textsc{WM-Basic}   & 0.4           & $(900, 3000)$   & $(-1.3, +0.3)$ & L10 \\
(TP-AGB)$^*$        &  .            & $(3500, 10^8)$ &                & L02 \\
\hline
\enddata
\tablenotetext{}{References: A12: \citet{2012RSPTA.370.2765A}, A14: \citet{2014IAUS..299..271A}, B96: \citet{1996ApJ...472..327B}, H19: \citet{2019AnA...621A..85H}, T15: \citet{2015AnA...579A..75T}, S12: \citet{2012AnA...540A.144S}, L10: \citet{2010ApJS..189..309L}, L02: \citet{2002AnA...393..167L}}
\tablenotetext{a}{Approximate wavelength sampling at $1000~\angstrom$}
\tablenotetext{b}{Approximate minimum/maximum wavelength sample}
\tablenotetext{c}{Approximate minimum/maximum metallicity sample}
\tablenotetext{*}{These are the default FSPS observational templates for TP-AGB stars, which have not been altered.}
\end{deluxetable}

\begin{figure}
    \centering
    \includegraphics[width=\columnwidth]{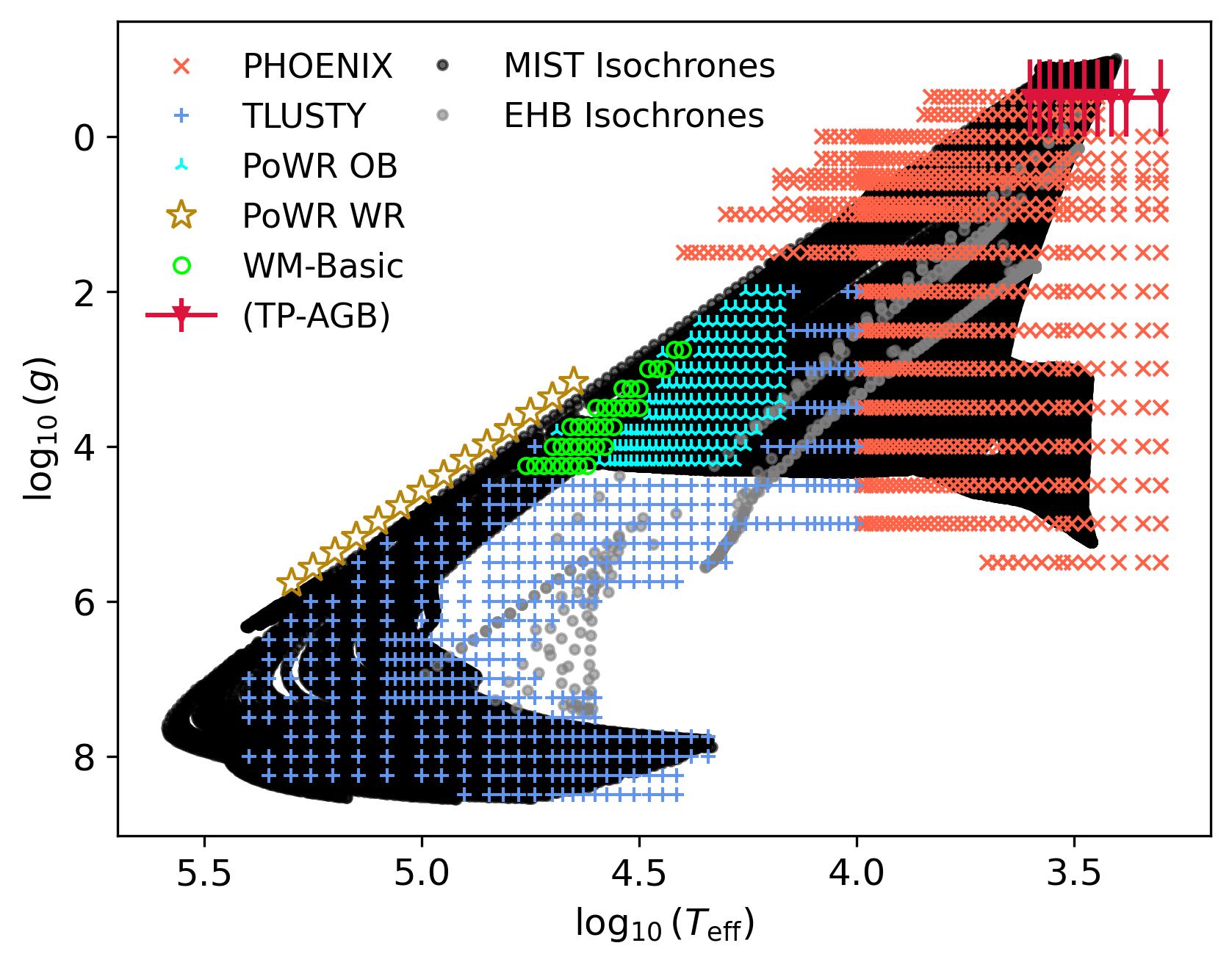}
    \caption{The coverage of our stellar template libraries in $\log T_{\rm eff}$-$\log g$ space is shown.  Note that both axes are reversed, such that they vary in the same way as a standard H-R diagram.  Isochrones from MIST tracks and EHB tracks are shown in the background for reference.}
    \label{fig:isochrones}
\end{figure}

For our analysis, some additional customizations outside of these default capabilities are required---in particular, the provided stellar template libraries in FSPS have a resolution of $10~\angstrom$ in the FUV, which is far too coarse to provide an adequate fit to our \textit{FUSE} spectra with a binned resolution of $0.2~\angstrom$.  We have therefore replaced FSPS's built-in stellar templates with a suite of high-resolution alternatives.  Due to the limited availability of high-resolution FUV observational templates, we use spectra derived from theoretical model atmospheres.  We follow the philosophy of FSPS by adopting different libraries for different parts of the H-R diagram, where the relative importances of different physical processes shift.  In the following text, we will outline the specifics of each newly adopted library.  The coverage of each library in $\log{T_{\rm eff}}-\log{g}$ space is shown in Figure \ref{fig:isochrones}, while the coverage in wavelength and $\log (Z/Z_\odot)$ is presented in Table \ref{tab:templatelibs}.  Note that stellar template libraries pulled from different sources may assume different solar metallicity and abundance standards.  FSPS can handle the difference in metallicity by explicitly providing the solar metallicity standard assumed by each template, and rescaling them to be on the same standard \citep[which we choose to be that of ][for which $Z_\odot = 0.0134$]{2009ARAnA..47..481A}.  However, the same standardization is not done for differences in individual elemental abundances, which will produce a systematic uncertainty in our metallicity measurements.

Pre-main sequence stars, cool main sequence stars ($T_{\rm eff} < 10,000$ K), and giant stars ($\log g < 2$) are covered by the BT-Settl grid \citep{2012RSPTA.370.2765A,2014IAUS..299..271A}\footnote{Retrieved from the Spanish Virtual Observatory at \url{https://svo2.cab.inta-csic.es/theory/newov2/}}.  These are calculated with the \textsc{Phoenix} code \citep{1999JCoAM.109...41H} with a static, spherically symmetric, 1D atmosphere model.  Importantly, the formation of molecules and dust clouds in the cool outer atmosphere layers are treated with physically motivated prescriptions.  While models exist at non-zero $\alpha$-enhancements, we consider only $[\alpha/{\rm Fe}] = 0$ models in this analysis.  We perform a 2D bilinear interpolation in $\log T_{\rm eff}$ and $\log g$ to place these templates on the same grid that FSPS assumes for its base stellar template library.  We do not extrapolate grid points outside the convex hull of the BT-Settl models, and instead set them to 0 so that they will be ignored during population synthesis. The wavelength sampling in the FUV is irregular, but typically ranges from $0.001$--$0.1~\angstrom$.  These models use abundances from \citet{2009ARAnA..47..481A}.

Hot O and B-type main sequence stars use a combination of grids from \citet{2019AnA...621A..85H} and \citet{2010ApJS..189..309L}. The \citet{2019AnA...621A..85H} grids use the Potsdam Wolf-Rayet ($\textsc{PoWR}$) code \citep{2015A&A...577A..13S}, while the \citet{2010ApJS..189..309L} grids use the \textsc{WM-Basic} code \citep{2001A&A...375..161P}.  In both cases, these codes model a spherically symmetric atmosphere with a stationary wind, while considering the effects of non-local thermodynamic equilibrium (NLTE), line blanketing, and wind inhomogeneities. The modeling of winds from hot stars is particularly important in the formation of the distinct P Cygni line profiles that are often seen in FUV stellar spectra, making this a primary consideration in our choice of stellar models in this parameter space.  The \textsc{PoWR} grids consist of a high-resolution UV spectrum (with a wavelength sampling of $\sim 0.01~\angstrom$ from $\sim 950$--$3000~\angstrom$), and a lower resolution SED (covering $\sim 5$--$10^9~\angstrom$), which we stitch together into a single spectrum.  The \textsc{WM-Basic} grids cover $\sim 900$--$3000~\angstrom$ at a sampling of $0.4~\angstrom$.  To cover the rest of the SED wavelength space, we stitch these templates together with the \textsc{Tlusty} grids (explained in the next paragraph) after interpolating them onto the same $\log T_{\rm eff}-\log g-\log Z$ grid.  The \textsc{PoWR} models use different abundance sets for each metallicity, appropriate for the Galactic \citep{2009ARAnA..47..481A}, LMC \citep{2007A&A...466..277H}, and SMC \citep{2007A&A...471..625T}, while the \textsc{WM-Basic} models use \citet{2005ASPC..336...25A}.

The remaining parameter space above $T_{\rm eff} > 10,000$ K is populated by post-main sequence stars and white dwarfs, which we cover with grids from \citet{1996ApJ...472..327B}.  These atmospheres are modeled with the \textsc{Tlusty} code \citep{1988CoPhC..52..103H}, which considers a static plane-parallel atmosphere in NLTE.  The original stellar templates computed by \citet{1996ApJ...472..327B} are smoothed to a FWHM resolution of $3~\angstrom$ and cover $900$--$1800~\angstrom$.  To improve the sampling and wavelength coverage, instead of using these spectra directly, we take the stellar atmosphere models and feed them into $\textsc{Synspec}$, the partner program to \textsc{Tlusty} for computing spectra \citep{2011ascl.soft09022H}, to recompute them from scratch.  The resultant spectra have a new sampling of $\sim 0.1~\angstrom$ in the UV and cover $90$--$10^6~\angstrom$.  We also must make a correction to the normalization of these spectra, as pointed out by \citet{1996ApJ...472..327B}, due to the exclusion of elements heavier than He in the structure of the model atmosphere (they are only added during the computation of the spectrum with \textsc{Sysnpec}).  The effective temperature is defined as the temperature at which the Stefan-Boltzmann law holds---therefore, we renormalize the templates such that the bolometric flux at the stellar surface equals $F_{\rm bol} \equiv \sigma_{\rm SB}T_{\rm eff}^4$ (where $\sigma_{\rm SB}$ is the Stefan-Boltzmann constant).  These templates use the solar abundance standard of \citet{1998SSRv...85..161G}. 

Wolf-Rayet (WR) stars, populating the line in $\log T_{\rm eff}$--$\log g$ space that denotes the Eddington limit, are treated as special cases in FSPS, and receive their own template spectra.  For these, we use templates from \citet{2015AnA...579A..75T} for nitrogen-rich WRs and \citet{2012AnA...540A.144S} for carbon-rich WRs.  These both use the \textsc{PoWR} model atmosphere code, explained above for the O/B star grids.  In a similar fashion, they provide high resolution ($\sim 0.1~\angstrom$) UV spectra from $900$--$3000~\angstrom$, and a lower resolution SED covering $\sim 5$--$10^9~\angstrom$, which we stitch together into a single spectrum.

Finally, thermally pulsing asymptotic giant branch (TP-AGB) stars are also treated as special cases in FSPS.  However, due to the difficultly of simulating the dynamic nature of these types of stars, in addition to the fact that their effective temperatures are low enough that they should not contribute significantly to the FUV continuum, we have decided to retain the default FSPS templates for these stars, which are average observational templates from \citet{2002AnA...393..167L} that cover a wavelength range of $3500$--$10^8~\angstrom$.

Once all of the stellar template libraries are collected, we resample them onto a common wavelength grid with the following sampling:
\begin{equation}
\begin{cases}
    90-800~\angstrom    & \lambda/\Delta\lambda = 20 \\
    800-1800~\angstrom  & \Delta\lambda = 0.2~\angstrom \\
    1800-9000~\angstrom & \Delta\lambda = 20~\angstrom \\
    9000-10^7~\angstrom & \lambda/\Delta\lambda = 20
\end{cases}
\end{equation}
This provides us with full SED coverage and adequate resolution in the wavelengths relevant for our \textit{FUSE} spectra, while maintaining an array size that still allows for rapid calculations. 

Our libraries also provide complete coverage in $\log T_{\rm eff}$--$\log g$ space, fully encompassing the MESA Isochrones and Stellar Tracks \citep[MIST;][]{2016ApJS..222....8D, 2016ApJ...823..102C, 2011ApJS..192....3P, 2013ApJS..208....4P, 2015ApJS..220...15P, 2018ApJS..234...34P} that FSPS uses, as well as some additional parameter space for hot, dense stars that is expected to be the home of so-called ``extreme horizontal branch'' (EHB) stars (see Figure \ref{fig:isochrones}).  These EHB tracks arise from stars with low envelope masses at the zero-age horizontal branch, which prevents them from developing the extensive outer convection zone seen in AGB stars, and forces them to stay at a high effective temperature ($\gtrsim 20,000$ K) throughout the rest of their evolution \citep{1993ApJ...419..596D}.  The origins of EHB stars are not well understood, as it is difficult to evolve such low envelope masses without significant mass loss.  Recent studies have found a link between EHB stars and enhanced multiplicity \citep{2025A&A...702A..11G}, suggesting that binary mass transfer plays a key role in their evolution.  Due to the extended period of time they spend at high $T_{\rm eff}$ relative to ``normal'' horizontal branch stars, EHB stars and their progeny are widely considered to be the origin of the ``UV upturn'' phenomenon seen in elliptical galaxies with little to no ongoing star formation \citep{1993ApJ...419..596D, 1997ApJ...482..685B, 2000ApJ...532..308B}.  By default, FSPS has a built-in option to include a fraction of EHB stars by modifying the MIST isochrones and manually placing new stars at evenly spaced intervals in $T_{\rm eff}$ up to 30,000 K.  We have instead opted for a more physically motivated approach by taking evolutionary tracks computed specifically for EHB stars by \citet{1993ApJ...419..596D}, interpolating onto the same time steps as the MIST isochrones, and including them directly in FSPS's \texttt{mod\_hb} routine.

\subsection{SED Models} \label{sec:continuum}

Before modeling the SED with stellar populations, we apply a few additional modifications to the input data.  Since we do not attempt to model line features (excluding those originating from stars), we mask the \textit{FUSE} spectra at the positions of known airglow lines from the Lyman series and \ion{O}{6} $\lambda\lambda$1031.9,1037.6 doublet within $\pm 180$ \kms~ (3 pixels); Galactic ISM absorption lines from \ion{O}{1} ($\lambda$988.6, $\lambda$988.7, $\lambda$988.8, $\lambda$1039.2), \ion{N}{3} ($\lambda$989.8, $\lambda$991.5, $\lambda$991.6), \ion{C}{2} $\lambda$1036.3, \ion{Ar}{1} $\lambda$1048.2, \ion{N}{2} $\lambda$1084.0, \ion{N}{1} ($\lambda$1134.2, $\lambda$1134.4, $\lambda$1135.0), \ion{Ca}{2} ($\lambda$1135.5, $\lambda$1135.6), and \ion{Fe}{2} ($\lambda$1142.4, $\lambda$1143.2, $\lambda$1144.9) within $\pm 60$ \kms~ (1 pixel); and at the redshifted positions of commonly detected extragalactic emission lines of \ion{C}{3} $\lambda$977.0, \ion{N}{3} ($\lambda$989.8, $\lambda$991.5, $\lambda$991.6), \ion{O}{6} $\lambda\lambda$1031.9,1037.6, and \ion{Ar}{1} $\lambda$1066.7 within $\pm 180$ \kms.  We mask out all wavelengths between 1082.7--1086.8 $\angstrom$ since this range falls within a gap between the coverage of \textit{FUSE}'s LiF detectors, leaving typically much lower SNR data from the SiC detectors. We also mask out the edges of the spectrum at $< 925\,\angstrom$ and $> 1175\,\angstrom$, since these are typically dominated by backgrounds.  We apply a correction to remove the reddening due to foreground Galactic dust along the line of sight to the objects of interest.  We obtain $E(B-V)$ values from the \citet{2011ApJ...737..103S} dust maps (via python's \texttt{astroquery.irsa\_dust} package), and de-redden the spectra and photometry using a \citet{1989ApJ...345..245C} extinction law.  To allow for the calculation of synthetic photometry when comparing models to observations, we obtain filter transmission curves for the relevant photometric bands from the Spanish Virtual Observatory's filter profile service\footnote{\url{http://svo2.cab.inta-csic.es/theory/fps/}}.

Using our updated FSPS backend (which is publically available through forked github repositories for both FSPS\footnote{\url{https://github.com/Michael-Reefe/fsps}} and Python-FSPS\footnote{\url{https://github.com/Michael-Reefe/python-fsps}}), we construct a \textsc{Prospector} model which consists of two stellar populations (which we refer to as the ``old'' and ``young'' populations), each with an exponential star formation history (SFH $\propto e^{-t/\tau}$), Salpeter IMF \citep{1955ApJ...121..161S}, and Calzetti attenuation curve \citep{2000ApJ...533..682C}.  The number of EHB stars in the isochrones are also allowed to vary, parametrized with the EHB fraction $f_{\rm EHB}$.  This affects both stellar populations, but since most stars in the young population won't have evolved to the horizontal branch, it is primarily relevant for the old population.  In addition to the stellar continuum, we include a variety of physically motivated components, some multiplicative and some additive, that represent other astrophysical and instrumental contributions to the spectrum.  

We model infrared dust emission (which must energetically balance the dust absorption) based on \citet{2007ApJ...657..810D}.  An optional AGN component with a ``clumpy torus'', as described by \citet{2008ApJ...685..147N}, is also included.  We model H$_2$ absorption as a function of wavelength based on \citet{2003PASP..115..651M}.  Allowing each individual H$_2$ transition to have an independent column density would lead to too many free parameters to marginalize over, so we instead parametrize in terms of the total H$_2$ column density, and set each energy level's columns using a continuous temperature distribution such that ${\rm d}N(T) \propto T^{-4.8}{\rm d}T$, with $T$ integrated from $49$--$2000$ K, following \citet{2016ApJ...830...18T}.  H$_2$ absorption is applied both for foreground Galactic H$_2$ along the line of sight, and for intrinsic H$_2$ within the target object.  Finally, we include instrumental calibration components, with the \textit{FUSE} background spectrum as an additive component with a free normalization, and a 4th-order multiplicative Chebyshev polynomial to correct for non-uniform absolute flux calibration.

A full summary of the model parameters, and the priors we place on them, can be found in Table \ref{tab:parameters}.  Our general approach is to be as conservative as possible when placing prior constraints on parameters, opting for uniform or log-uniform priors in nearly all cases.  The only parameter we constrain with a stronger prior is the AGN fraction $f_{\rm AGN}$, for which we impose a clipped normal prior centered at $\mu=0$ and with a dispersion of $\sigma=0.03$, clipped to the range $(0,1)$.  This corresponds to a requirement that measuring an AGN fraction of $\sim 10$\% requires a $\sim 3\sigma$ deviation in the prior.  This is done to follow Occam's razor---preventing an AGN component from being included to model the infrared emission unless it cannot be solely explained by stellar-reprocessed dust emission.

The choice to include $f_{\rm EHB}$ as a free parameter was made after finding that it was able to produce marginally better models for high-SNR spectra.  However, performing a statistical $F$-test reveals that these models are, in general, not consistently preferred at the $3\sigma$ level over models where $f_{\rm EHB}$ is fixed at 0.  And in the case of low-SNR spectra, the models are nearly indistinguishable to any significant criterion.  The choice for whether to include $f_{\rm EHB}$ can have a significant effect on the posteriors for other model parameters---in particular SFRs.  So, to study the systematic uncertainty that this choice imposes on our results, we run another set of models with $f_{\rm EHB} = 0$ for comparison purposes.

\begin{deluxetable}{llCclll}
\def\arraystretch{0.85}
\tabletypesize{\footnotesize}
\tablecaption{\textsc{Prospector} model configuration}
\label{tab:parameters}
\tablehead{
    \colhead{Parameter} & \colhead{Unit} & \colhead{Initial Value} & \colhead{Prior} & \colhead{Notes} & \colhead{Ref.}
}
\decimals
\startdata
\hline
SFH type        &            & 1       & \faLock                         & Exponential SFH      & J21 \\
IMF type        &            & 0       & \faLock                         & Salpeter IMF         & S55 \\
Dust type       &            & 2       & \faLock                         & Calzetti dust attenuation curve & C00  \\
$z$             &            & $z_0$   & \faLock                         & Redshift (from Table \ref{tab:info}) & \\
$D$           & Mpc        & $D_0$   & \faLock                         & Distance (from Table \ref{tab:info}) & \\
$M_{*,\rm old}$ & $M_\odot$  & 10^{10} & $\mathcal{L}(10^{6}, 10^{14})$  & Old population mass  & J21 \\
$M_{*,\rm yng}$ & $M_\odot$  & 10^{8}  & $\mathcal{L}(10^{6}, 10^{10})$  & Young population mass ($<$ old pop mass) & J21 \\
$\log(Z_{*,\rm old}/Z_\odot)$ && -1.0    & $\mathcal{U}(-1.3, +0.0)$      & Old population metallicity & J21 \\
$\log(Z_{*,\rm yng}/Z_\odot)$ && +0.0    & $\mathcal{U}(-1.3, +0.0)$      & Young population metallicity & J21 \\
$t_{\rm age,old}$ & Gyr       & $t_0$\tablenotemark{a} & $\mathcal{U}(t_0, t_{\rm univ}(z))$   & Old population age & J21 \\
$t_{\rm age,yng}$ & Gyr       & 10^{-3} & $\mathcal{U}(0,1)$      & Young population age & J21 \\
$\tau_{\rm old}$  & Gyr       & 0.1     & $\mathcal{L}(10^{-3},1)$          & Old population e-folding time & J21 \\
$\tau_{\rm yng}$  & Gyr       & 0.1     & $\mathcal{L}(10^{-3},100)$      & Young population e-folding time & J21 \\
$\tilde{\tau}_{\rm V,old}$  && 0.5     & $\mathcal{U}(0,4)$              & Old population optical depth at $5500\,\angstrom$ & J21 \\
$\tilde{\tau}_{\rm V,yng}$  && 1.0     & $\mathcal{U}(0,4)$              & Young population optical depth at $5500\,\angstrom$ & J21 \\
$f_{\rm EHB}$               && 0.0     & $\mathcal{U}(0.0,0.5)$          & Fraction of EHB stars : normal HB stars & C09 \\
EHB comp.                   && 8       & \faLock                         & Composition of EHB stars (see set ``H'' in B97) & B97 \\
$s_{\rm BS}$                && 0.0     & \faLock                         & Specific frequency of blue stragglers & C09 \\
$\Delta \log L_{\rm bol}$   && 0.0     & \faLock                         & Shift in $\log L_{\rm bol}$ of TP-AGB isochrones & C09 \\
$\Delta \log T_{\rm eff}$   && 0.0     & \faLock                         & Shift in $\log T_{\rm eff}$ of TP-AGB isochrones & C09 \\
$U_{\rm min}$ & $u_{\rm M83}$\,\tablenotemark{b} & 1.0     & $\mathcal{U}(0.1, 25)$          & Minimum intensity of dust-heating radiation & D07 \\
$q_{\rm PAH}$ & \%          & 4.0     & $\mathcal{U}(0.5, 7)$           & Mass fraction of PAHs : dust & D07 \\
$\gamma$      &             & 10^{-3} & $\mathcal{L}(10^{-3},0.15)$     & Mass fraction of dust heated by $U > U_{\rm min}$ & D07 \\
$f_{\rm AGN}$ & $L_{\rm AGN}/L_*$ & 10^{-4} & $\mathcal{C}(0,0.03;0,1)$       & AGN luminosity fraction & J21 \\
$\tilde{\tau}_{\rm AGN}$    && 5       & $\mathcal{L}(5,150)$            & AGN torus optical depth & J21 \\
$\sigma_*$ & \kms           & 100     & $\mathcal{U}(15,700)$           & Velocity dispersion & J21 \\
$n_{\rm bkg}$               && 1.0     & $\mathcal{U}(0.7,1.3)$          & \textit{FUSE} background spectrum normalization & \\
$P_{\rm order}$             && 4       & \faLock                         & Multiplicative Chebyshev polynomial order & J21 \\
$P_{\rm reg}$               && 0       & \faLock                         & Polynomial regularization & J21 \\
$\log (N_{\rm H_2,fg}/{\rm cm^{2}})$ && $\log (N_{\rm H_{2},0})$\tablenotemark{c} & $\mathcal{U}(15,21)$           & Log of H2 column density (Milky Way) & \\
$\log (N_{\rm H_2,in}/{\rm cm^{2}})$ && $\log (N_{\rm H_{2},0})$\tablenotemark{c} & $\mathcal{U}(15,21)$           & Log of H2 column density (intrinsic) & \\
\hline
\enddata
\tablenotetext{}{Priors: $\mathcal{U}(m,n)$ denotes a uniform prior over the range $(m,n)$.  $\mathcal{L}(m,n)$ denotes a log-uniform prior over the range $(m,n)$.  $\mathcal{C}(\mu,\sigma;m,n)$ denotes a clipped normal prior with a mean $\mu$ and dispersion $\sigma$, clipped to the range $(m,n)$. \faLock~ denotes that the parameter is locked.}
\tablenotetext{}{Listed references contain explanations and/or definitions for the relevant parameter. J21: \citet{2021ApJS..254...22J}, S55: \citet{1955ApJ...121..161S}, C00: \citet{2000ApJ...533..682C}, C09: \citet{2009ApJ...699..486C}, B97: \citet{1997ApJ...482..685B}, D07: \citet{2007ApJ...657..810D}}
\tablenotetext{a}{Initial old population age is estimated from the formula $t_0 =t_{\rm univ}(z)-t_{\rm univ}(3)$ where $t_{\rm univ}(z)$ represents the age of the universe at redshift $z$.}
\tablenotetext{b}{In units normalized to the Milky Way interstellar radiation field calculated by \citet{1983AnA...128..212M}}
\tablenotetext{c}{Initial H2 column densities are estimated from the formula $N_{\rm H_{2,0}} \approx (f_{\rm H_2}/2)\,5.8 \times 10^{21}E(B-V)$ (REFS).}
\end{deluxetable}

\subsection{X-ray Spectrum Models} \label{sec:model_xspec}

Parallel to our UV--IR SED models, we perform a separate analysis of the X-ray spectra with \textsc{PyXspec} \citep{2021ascl.soft01014G}, with the primary goal of obtaining metallicities and X-ray luminosities for the hot phase.  We fit each spectrum from 0.5--7 keV with both a 1-temperature \texttt{phabs(apec)} model and a 2-temperature \texttt{phabs(apec+apec)} model, fixing the redshifts to the values in Table \ref{tab:info}.  We subtract backgrounds before fitting as described in $\S$\ref{sec:chandra}---we opt out of any more sophisticated analysis since the aperture considered is small and centered on the highest $S/N$ part of the system.  For central cluster galaxies with plasma temperatures $\gtrsim 4$ keV, we mask out energies from 0.9--1.3 keV following \citet{2021A&A...646A..92G} to avoid systematic biases that have been observed in the Fe L-shell region.  We use Cache statistics (\texttt{cstat}) as our optimization criterion.

The CGM abundance values retrieved from these fits are converted from the default \citet{1989GeCoA..53..197A} units to \citep{2009ARAnA..47..481A} units, so that they are comparable to the stellar metallicities measured from our SED models.

\section{Results} \label{sec:results}

\begin{figure*}
    \centering
    \includegraphics[width=\textwidth]{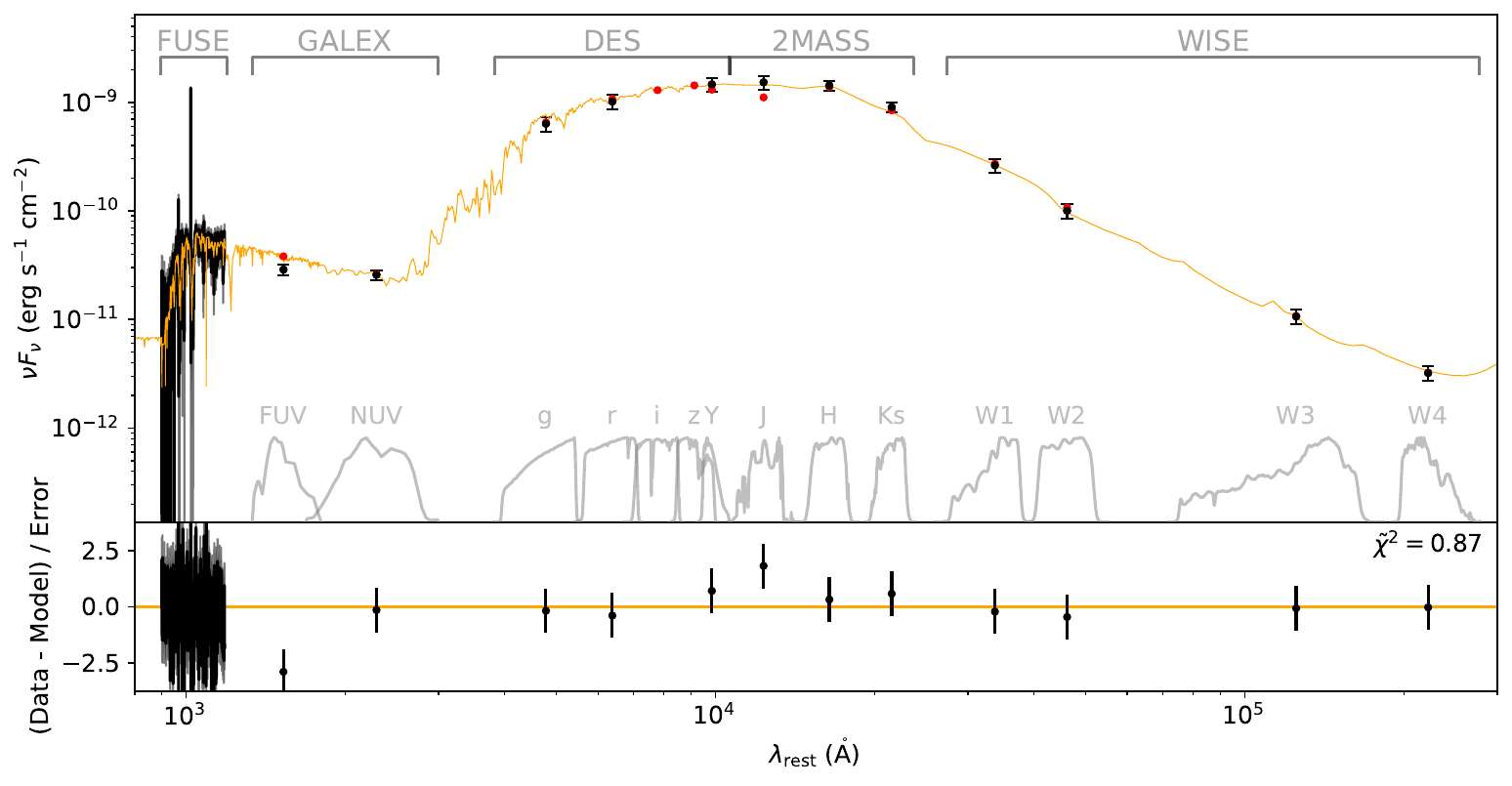}
    \includegraphics[width=\textwidth]{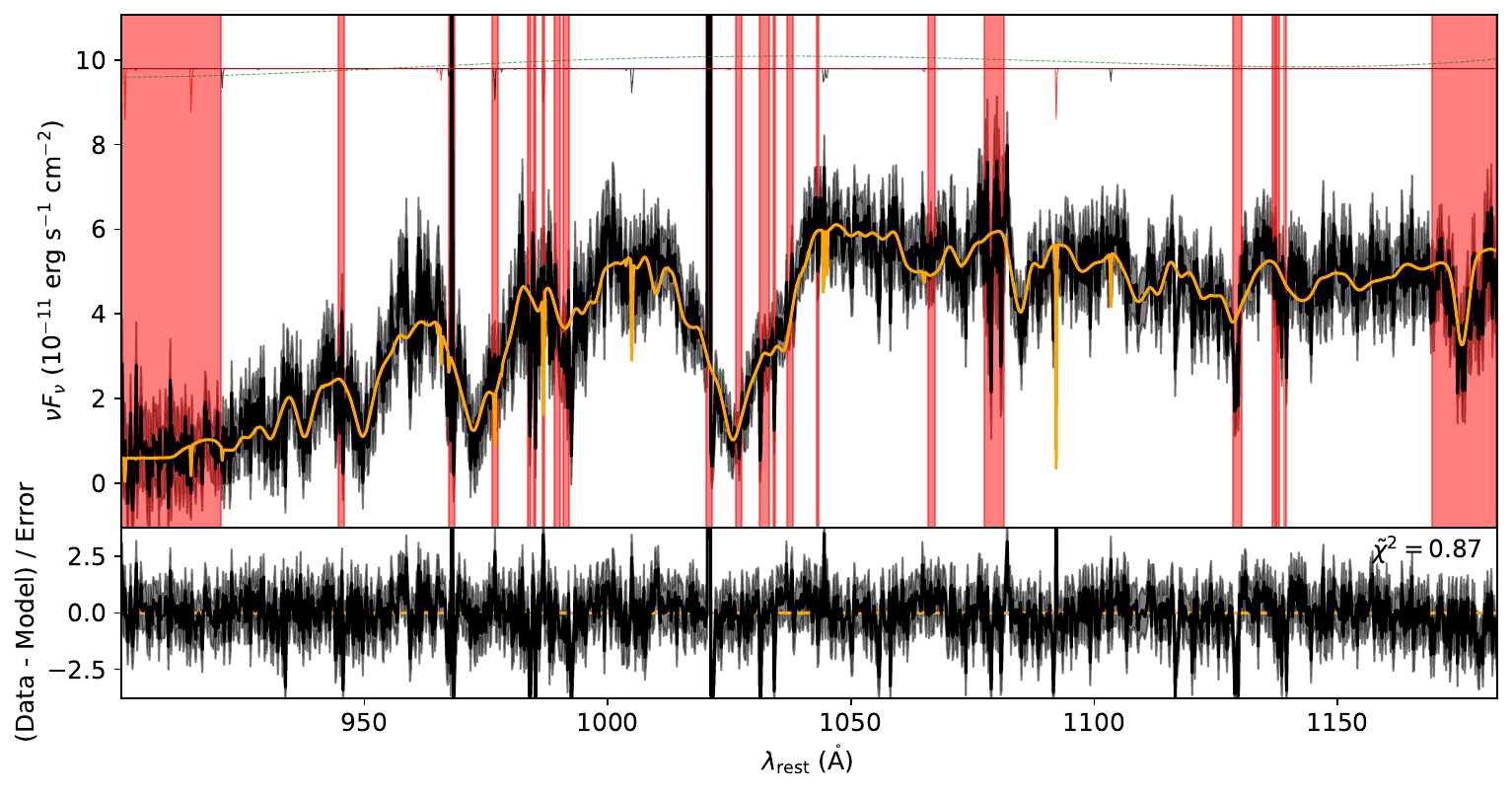}
    \caption{The maximum a posteriori model for Fornax (NGC 1399).  \textit{Top panel}: The full SED is shown, with data from \textit{FUSE}, \textit{GALEX}, DES, 2MASS, and \textit{WISE} labeled accordingly (note that the DES i and z band data is missing due to saturation).  The data and errors are in black, and the model is in orange.  The transmission functions for each broadband filter are shown in gray at the bottom of the plot. Below this, the relative residuals, defined as $({\rm Data}-{\rm Model})/{\rm Error}$, are shown.  The reduced chi-squared statistic ($\tilde{\chi}^2$) is given in the top-right corner of the residuals plot. \textit{Bottom panel}: A zoomed-in view of the \textit{FUSE} spectrum is shown.  The data is in black, with the errors shown by the gray region, and the model is in orange.  Translucent red bands shown parts of the spectrum which have been masked out.  Some additional information is shown at the top of the plot: The thin black line shows the foreground Galactic H$_2$ absorption (which is negligible for this example), the thin red line shows intrinsic H$_2$ absorption (which is also almost negligible), and the thin dashed green line shows the shape of the 4th-order multiplicative polynomial.  The scale of these lines (not shown on the plot) is normalized such that the top of the plot is 2, and the bottom of the deepest H$_2$ absorption feature is 0.  The relative residuals and $\tilde{\chi}^2$ are also shown at the bottom of the plot.}
    \label{fig:fornax_sed}
\end{figure*}

\begin{figure}
    \centering
    \includegraphics[width=\columnwidth]{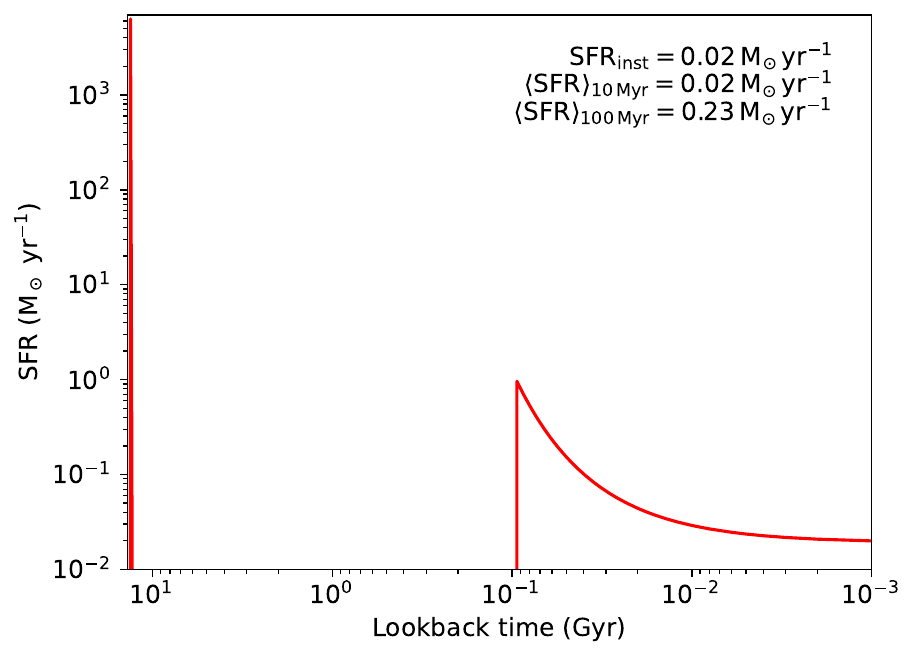}
    \caption{The SFH of the maximum a posteriori model for Fornax (NGC 1399).  A few different SFRs, averaged over different time intervals (instantaneous, 10 Myr, and 100 Myr), are annotated in the top-right of the plot.}
    \label{fig:fornax_sfh}
\end{figure}

We show an example of a maximum a posteriori SED model of the Fornax BCG (NGC 1399) in Figure \ref{fig:fornax_sed}.  We choose this system as an example because of its high signal-to-noise ratio and relative lack of prominent nebular emission/absorption lines, which has led it to historically be used as a template spectrum for modeling the stellar continuum in other elliptical galaxies.  Our model provides a statistically robust fit to the data while avoiding overfitting, with a reduced chi-squared statistic $\tilde{\chi}^2 = 0.91$.  This is generally representative of the fit quality in all 30 systems.

We show the underlying model's SFH in Figure \ref{fig:fornax_sfh}.  The shape of this SFH at ages older than $\sim 1$ Gyr is not well constrained by the data we have---our photometry provides solid constraints on the integrated properties of the old population (age, mass, metallicity, extinction), but the time-dependent properties (SFH shape) are not as constrained and would require optical spectroscopy.
Therefore, one should be cautious when drawing any conclusions from the ``old'' portion of this SFH, which is shown primarily for illustrative purposes. The ``young'' portion $<1$ Gyr, meanwhile, is better constrained by \textit{FUSE}, for which we find a 100 Myr averaged SFR of $0.21 \pm 0.03$ \msunyr.  This means that a genuinely young stellar population accounts for roughly $77 \pm 5$\% of the FUV flux integrated between 900--1200 \angstrom, with the other 22\% being from EHB stars, for which we find an EHB fraction of $f_{\rm EHB} = 0.02 \pm 0.01$ (recall that $f_{\rm EHB}$ is defined as the number ratio of EHB stars to normal HB stars).

This is somewhat in tension with previous analyses of this system, which have suggested that the FUV flux must be at least 50\% a result of EHB stars due to a lack of detectable \ion{C}{4} $\lambda$1550 absorption \citep{1991ApJ...382L..69F, 1997ApJ...482..685B}.  However, we find that by including a higher fraction of young stars, the model better represents the observed continuum between $\sim 1030$--$1050$ \angstrom, which is shaped in part by P Cygni profiles from the \ovi\, doublet, arising from strong winds in O and B stars.  Additionally, it allows the old stellar population to explore a higher metallicity, which better fits the broadband optical--IR SED data.  

We note, however, an important caveat---since we use synthetic spectra libraries from different codes with different assumptions for separate regions of the HR diagram, this could drive a systematic bias towards young stars or EHB stars in our models, which is difficult to quantify.  Nevertheless, no single library, synthetic or observational, exists that covers an adequate parameter space to solely use for our models.  In practice, we find that many of our sources, being at lower $S/N$ than NGC 1399, have much less confident decompositions between young and EHB stars, which is likely reflective of the true physical uncertainties given the data, whereas the systems with more confident young stellar components are the ones where such a contribution is expected (due to known strong cooling flows in e.g. Perseus, Abell 1795).

In the next few subsections, we will review the population-wide results for a subset of the parameters.

\subsection{Star formation rates and EHB fractions} \label{sec:results_sfr}

\begin{figure}
    \centering
    \includegraphics[width=\columnwidth]{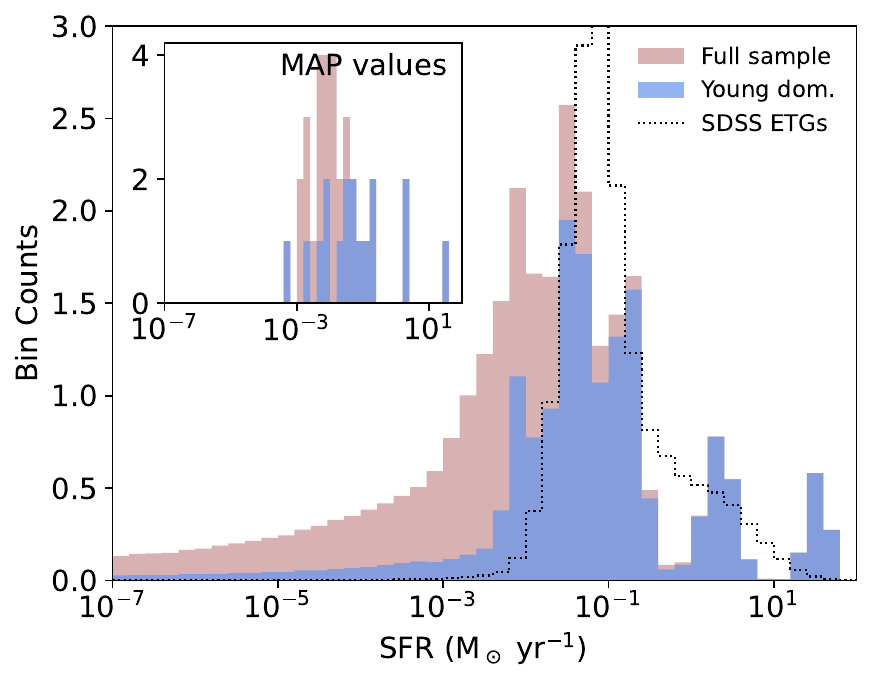}
    \caption{Histograms of SFRs for our full sample (red), young-dominated systems with $f_{\rm young} \geqslant 0.1$ (blue), and a comparison sample of SDSS early type galaxies from the MPA/JHU catalogue (dashed black line).  The main panel shows the cumulative posterior distributions, while the inset panel shows the discretized histograms of the maximum a posteriori (MAP) values of each posterior.}
    \label{fig:sfr_posterior}
\end{figure}

\begin{figure}
    \centering
    \includegraphics[width=\columnwidth]{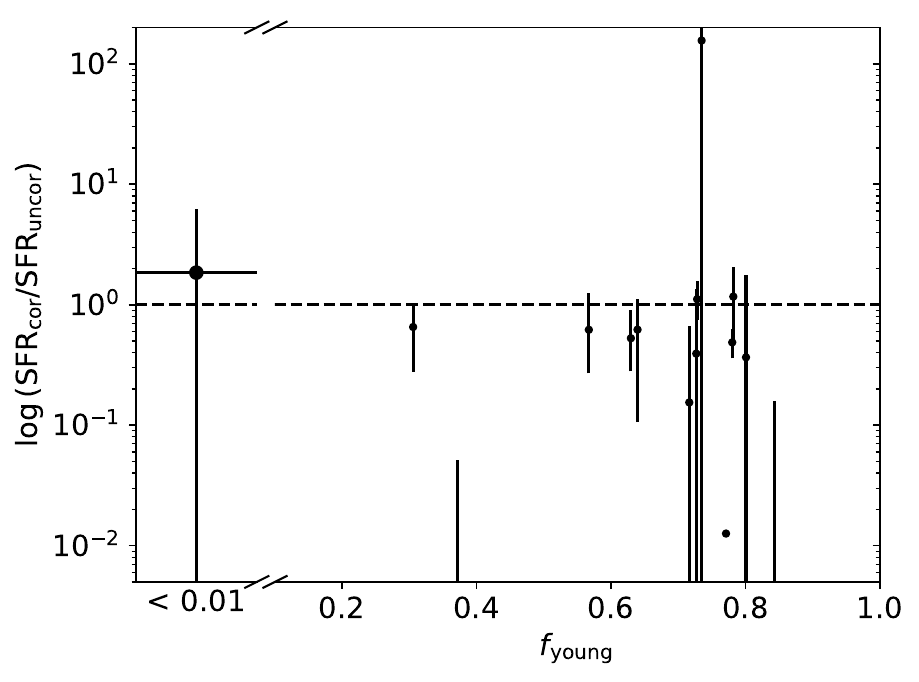}
    \caption{The fraction of the ``corrected'' ${\rm SFR}_{\rm cor}$, measured when $f_{\rm EHB}$ is allowed to vary, and the ``uncorrected'' ${\rm SFR}_{\rm uncor}$, measured when $f_{\rm EHB}$ is fixed at 0.  Systems with $f_{\rm young} < 0.01$ are grouped and plotted as a single bin, where the error bars show the scatter.}
    \label{fig:sfr_compare}
\end{figure}

A histogram of SFRs is given in Figure \ref{fig:sfr_posterior}.  The main panel shows the cumulative posterior distribution of SFRs for all galaxies in the sample (in red), while the inset panel shows the discretized histogram of the maximum a posteriori values from each posterior.  The distribution peaks at an SFR of $\sim 0.03\,\msunyr$, but with extrema ranging from $\sim 10^{-5}$--$10^{1}$ \msunyr.

We compare this distribution with the SFRs measured in early-type galaxies in SDSS (dotted black line).  We obtain these SFR measurements from the Max Planck Institute for Astrophysics / Johns Hopkins University (MPA/JHU) catalogue\footnote{\url{https://www.sdss4.org/dr17/spectro/galaxy\_mpajhu/}}, which we then cross match with the Galaxy Zoo 2 project \citep{2008MNRAS.389.1179L, 2016MNRAS.461.3663H} using the SpecObjIDs to obtain Hubble classifications for each system.  We then filter the results to obtain SFRs only for galaxies classified as early types, which leaves us with 11,991 galaxies.  This distribution peaks at a higher SFR than our sample, at $\sim 0.08\,\msunyr$, and experiences a steep drop-off for SFRs $\lesssim 0.03\,\msunyr$.  This is likely due to SDSS's sensitivity limitations, as SFRs lower than this will leave vanishingly faint signatures in optical spectrophotometry.

In an attempt to mimic the SDSS sensitivity function, we also plot a histogram (in blue) of our sample's SFRs filtered to only sources which have at least 10\% of their integrated FUV light produced by a young stellar population.  We quantify this with the ``young fraction'' $f_{\rm young}$, defined as:
\begin{equation}
    f_{\rm young} = \frac{\int_{900\,\angstrom}^{1200\,\angstrom}L_{\lambda,\rm young}{\rm d}\lambda}{\int_{900\,\angstrom}^{1200\,\angstrom}[L_{\lambda,\rm young}+L_{\lambda,\rm old}]\,{\rm d}\lambda}
    \label{eq:fyoung}
\end{equation}
where $L_{\lambda,\rm young}$ is the specific luminosity of the young stellar population, and $L_{\lambda,\rm old}$ is the specific luminosity of the old stellar population.  Sources with $f_{\rm young} \geqslant 0.1$ are included in this histogram. We find this distribution to be a good match to the SDSS one.  A 2-sample Kolmogorov-Smirnov (KS) test between the two sample distributions gives a $p=0.82$, indicating that the two sample distributions are most likely drawn from the same underlying distribution. This supports the hypothesis that the difference in our full sample's distribution is sensitivity-driven, rather than population-driven.

The larger width in the distribution of SFRs in the full sample, and the long extended tail to SFRs $\ll 10^{-3}\,\msunyr$, is a result of a few factors.  A significant fraction of our sources have a faint FUV continuum, which makes the decomposition between young stars and EHB stars highly uncertain.  Because we allow $f_{\rm EHB}$ to vary, this results in wide, weakly constrained SFR posteriors for these sources.  Even for sources with brighter continua, uncertainties in this decomposition can result in large variations in the SFR posteriors.  In Figure \ref{fig:sfr_compare}, we show the fraction ${\rm SFR}_{\rm cor}/{\rm SFR}_{\rm uncor}$, where the ``corrected'' ${\rm SFR}_{\rm cor}$ is the (100 Myr averaged) SFR measured when $f_{\rm EHB}$ is allowed to vary, and the ``uncorrected'' ${\rm SFR}_{\rm uncor}$ is the SFR measured when $f_{\rm EHB}$ is fixed at 0.  The SFR fraction is $< 1$, or within $\pm 1\sigma$ of 1, for all systems---as expected if EHB stars are masquerading as additional star formation when they are not properly accounted for within the model.  This trend is tighter for systems with $f_{\rm young}>0.1$, as their SFRs are generally better constrained, whereas the ratio becomes much noisier at $f_{\rm young}<0.1$. This conclusion, while expected, should be taken with caution, as comparing the distributions of ${\rm SFR}_{\rm cor}$ and ${\rm SFR}_{\rm uncor}$ with a KS test reveals that the distributions are not significantly different from each other at more than a $\sim1\sigma$ level ($p = 0.10$) when only including the systems with the most constrained SFRs ($f_{\rm young}>0.1$). Nevertheless, the inclusion of $f_{\rm EHB}$ as a free parameter remains necessary for an accurate determination of the (systematic) uncertainty on the SFRs, obtained from the sampling posteriors.

\begin{figure}
    \centering
    \includegraphics[width=\columnwidth]{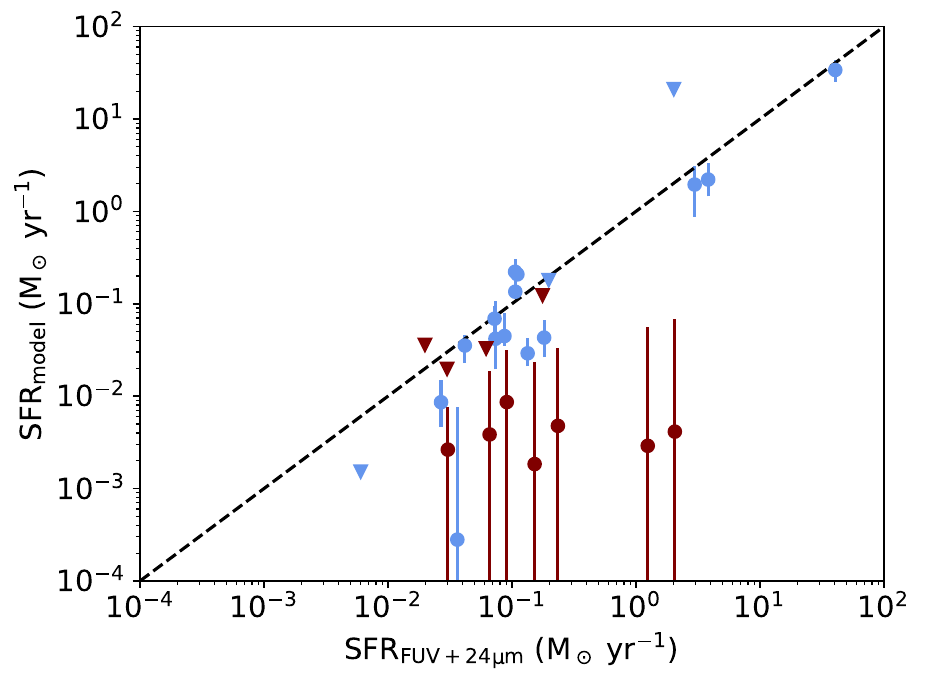}
    \caption{A comparison between the SFRs from our SED models (SFR$_{\rm model}$), and the SFRs inferred from a 2-band scaling relation using \textit{GALEX} FUV and \textit{WISE} W4 photometry (SFR$_{\rm FUV+24\mu m}$).  The dashed black line shows what a 1:1 correlation would look like.  The model SFRs are averaged over the past 100 Myr, matching the assumed duration of star formation from the 2-band photometry relation. The data is split into systems with $f_{\rm young} > 0.03$ (in blue) and $< 0.03$ (in red).}
    \label{fig:sfr_ltir}
\end{figure}

A similar effect is seen when comparing the SFRs to those that would be inferred using a purely photometric SFR scaling relation.  For comparison purposes, we use \textit{GALEX} FUV band luminosities, which have been corrected for dust extinction using the \textit{WISE} W4 band luminosities following \citet{2011ApJ...741..124H}.  The final scaling relation then follows the form $({\rm SFR}_{\rm FUV+24\mu m}/\msunyr) = 10^{-43.147} [(\nu L_\nu(FUV) + 3.89 \times \nu L_\nu(24{\rm \mu m}))/\ergs]$.  We compare this with our modeled SFRs after averaging over the past 100 Myr in Figure \ref{fig:sfr_ltir}. We observe 14/33 ($\sim40$\%) of our SFRs to be consistent with the ${\rm SFR}_{\rm FUV+24\mu m}$ within $2\sigma$, while most of the rest are smaller. 

This disagreement likely stems from a few sources, but we believe the primary driver is our model's inclusion of the UV upturn from EHB stars.  The ${\rm FUV}+24{\rm \mu m}$ photometric relation assumes that all of the FUV emission results from young star formation, and therefore it overestimates SFRs for systems which are dominated by an old UV upturn component.  This is exactly what we observe in Figure \ref{fig:sfr_ltir} when the modeled SFRs fall below the 1:1 line, highlighted by the differently colored points, which show the EHB-dominated systems ($f_{\rm young}<0.03$) in red and star-forming systems ($f_{\rm young}>0.03$) in blue.  The remaining differences in the SFRs likely stem from many smaller factors.  In some systems, the IR luminosity traced by \textit{WISE} may be elevated by sources unrelated to star formation, leading to an over-correction in the dust extinction.  For example, a significant contribution from an AGN is likely present in Perseus and M87, and may be present at smaller levels in other systems. 

\begin{figure}
    \centering
    \includegraphics[width=\columnwidth]{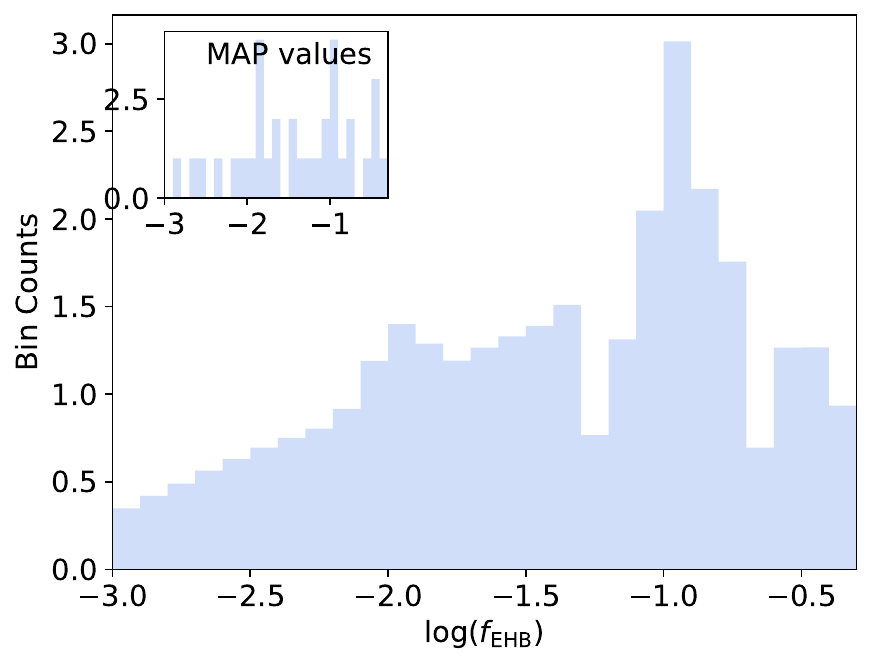}
    \caption{Histograms of the EHB fraction $f_{\rm EHB}$ for our full sample.  The main panel shows the cumulative posterior distribution, while the inset panel shows the discretized histogram of the maximum a posteiroriori (MAP) values.}
    \label{fig:fbhb_posterior}
\end{figure}

The distribution of $f_{\rm EHB}$ itself is shown in Figure \ref{fig:fbhb_posterior}.  One third of our sources cluster around $f_{\rm EHB} \sim 10^{-1}$, showing strong evidence for an EHB contribution.  Another $\sim$7 sources have $f_{\rm EHB} \lesssim 10^{-2}$, indicating little to no EHB contribution.  The rest fall somewhere in between these two cases.

\subsection{Ages} \label{sec:results_ages}

\begin{figure*}
    \centering
    \includegraphics[width=\columnwidth]{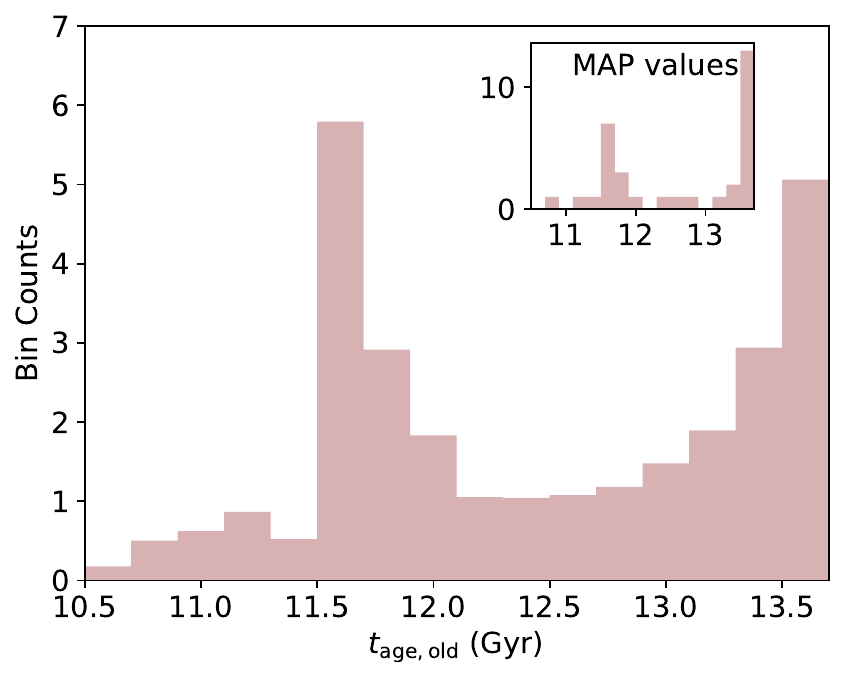}
    \includegraphics[width=\columnwidth]{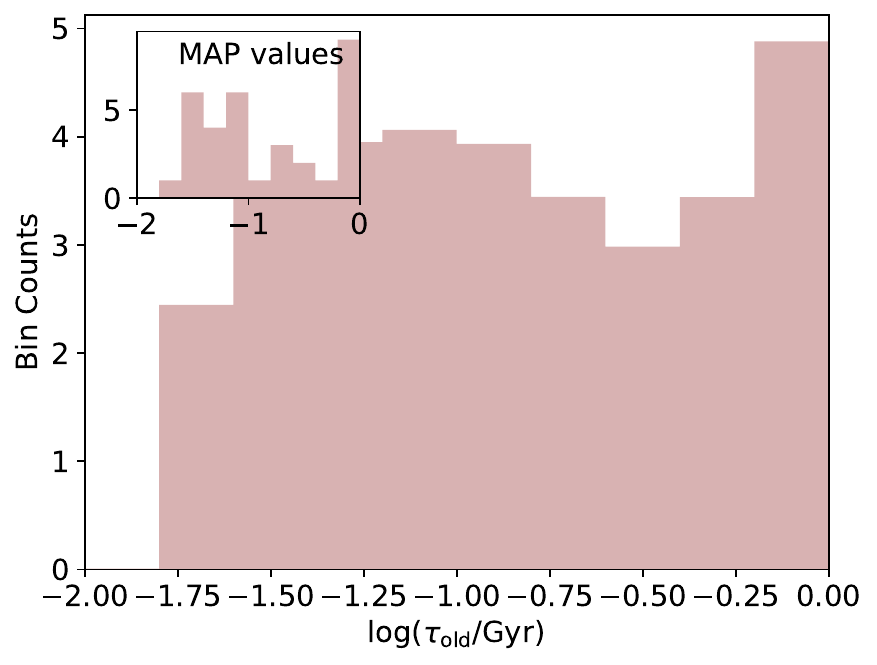}
    \caption{Histograms of the old stellar population's ages (left panel) and exponential decay timescales (right panel).  In each plot, the main panel shows the cumulative posterior distribution, while the inset panel shows the discretized histogram of maximum a posteriori (MAP) values.}
    \label{fig:t_old_posterior}
\end{figure*}

A histogram of old stellar population ages and decay timescales is shown in Figure \ref{fig:t_old_posterior}.  Like the SFR histogram, the main panel shows the cumulative posterior distribution, while the inset panel shows the discretized histogram of the maximum a posteriori values.  Since our photometric data do not have robust constraining power on the shape of the SFH for the old stellar population, our prior boundaries often impose stricter constraints on these parameters than the data themselves do, in an effort to keep our models internally consistent and interpretable.  For example, there is a noticeable artificial ``wall'' in the $t_{\rm age,old}$ distribution at 11.64 Gyr, which corresponds to the lookback time from redshift 0 to redshift 3, which is our imposed lower limit on $t_{\rm age,old}$ at $z=0$ (see Table \ref{tab:parameters}). This indicates mild evidence for some of these systems to have started forming later than $z=3$.  

Additionally, the $\tau_{\rm old}$ distribution shows a pile-up at the upper boundary of 1 Gyr.  This upper boundary was imposed to keep the ``old'' and ``young'' stellar populations strictly separate from each other in our models (as allowing for longer decay times could cause non-negligible star formation within the past 100 Myr to be attributed to the ``old'' population).  Nevertheless, the $\tau_{\rm old}$ posteriors suggests that there is some marginal evidence for some level of continuous star formation over the entire history of these galaxies.

\begin{figure*}
    \centering
    \includegraphics[width=\columnwidth]{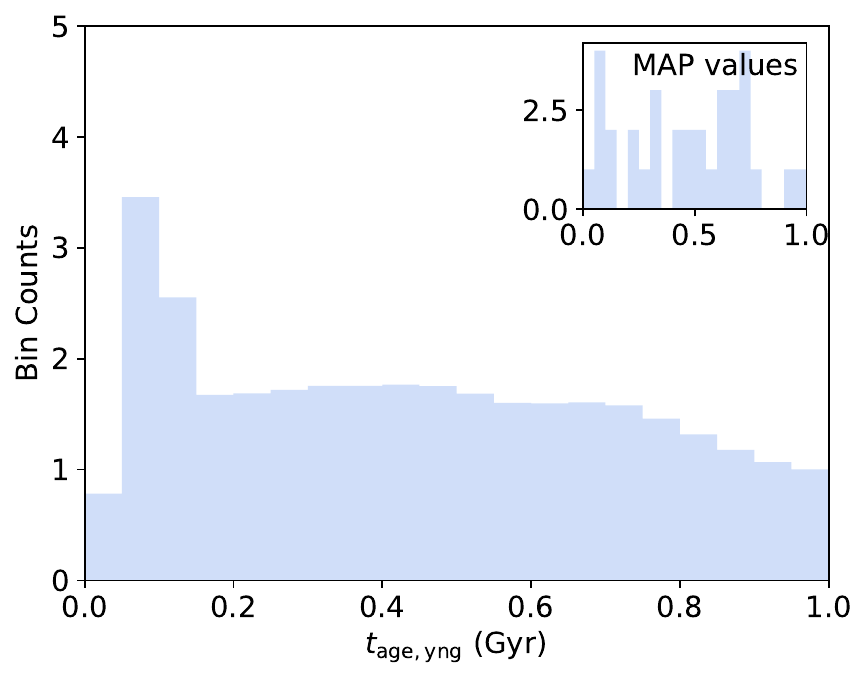}
    \includegraphics[width=\columnwidth]{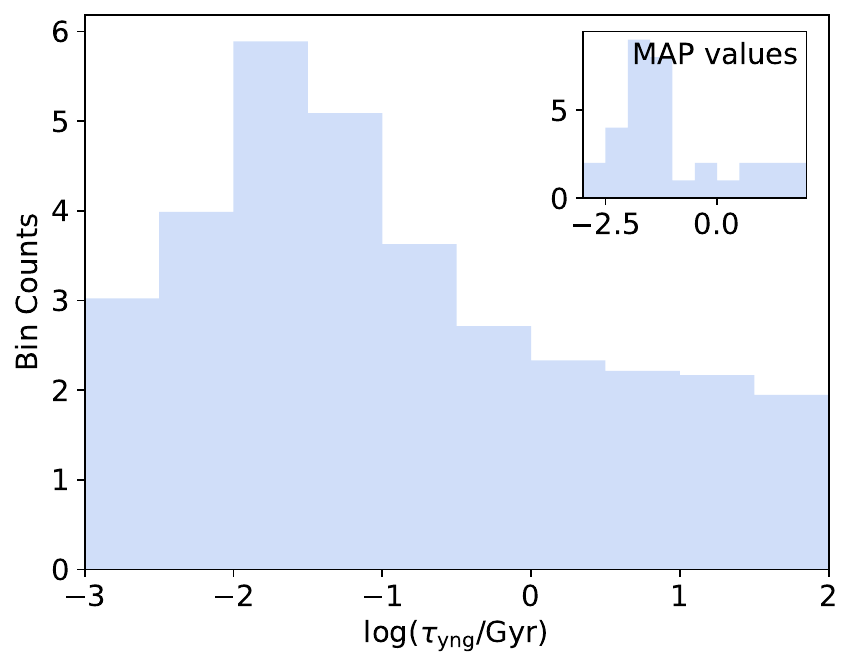}
    \caption{Equivalent to Figure \ref{fig:t_old_posterior}, but for the young stellar population.}
    \label{fig:t_young_posterior}
\end{figure*}

Corresponding age and decay time histograms for the young stellar population are shown in Figure \ref{fig:t_young_posterior}.  The $t_{\rm age,yng}$ posterior is relatively flat over the full range from 0--1 Gyr, showing that there is no strong preference in turn-on times for recent star formation.  There is an irregular peak between $\sim 50$--$150$ Myr, but interestingly, a similar peak is not seen in the maximum a posteriori histogram.  This means the peak is likely created by a small handful of systems which happen to have their $t_{\rm age,yng}$ posteriors constrained much tighter than the rest of the systems, rather than an overabundance of systems whose true $t_{\rm age,yng}$ lands in this range.

Decay times for the young population seem to prefer times $\lesssim 100$ Myr, as a similar trend is seen in both the cumulative posteriors and in the maximum a posteriori histogram.  In other words, recent star formation appears to be more ``bursty'' rather than continuous.

\subsection{Metallicities} \label{sec:results_logz}

\begin{figure}
    \centering
    \includegraphics[width=\columnwidth]{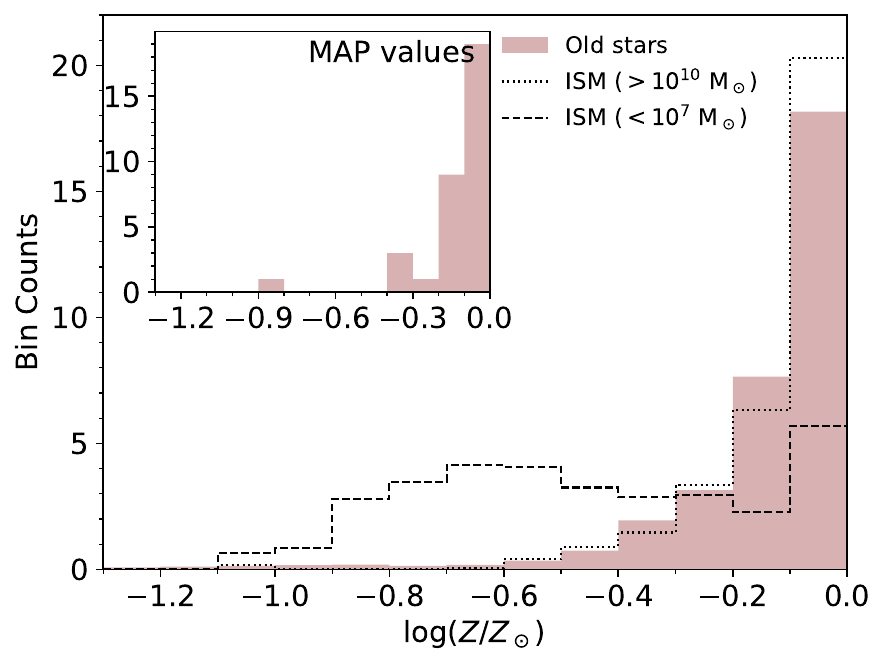}
    \caption{Histograms of the old stellar population's metallicity, in log solar units.  The main panel shows the cumulative posterior distribution, while the inset panel shows the discretized histogram of maximum a posteriori (MAP) values.  The dotted black line shows gas-phase ISM metallicities from SDSS galaxies with stellar masses $>10^{10}\,\msun$, while the dashed black line shows those with stellar masses $<10^{7}\,\msun$.}
    \label{fig:logzsol_old_posterior}
\end{figure}

The metallicity distribution of the old stellar population is shown in Figure \ref{fig:logzsol_old_posterior}.  We observe that the old stars are overwhelmingly metal enriched, with all but one source having $\log(Z_{*,\rm old}/Z_\odot) > -0.4$, and with many hitting our upper limit of $\log(Z_{*}/Z_\odot)=0$, indicating likely super-solar abundances.  For comparison, we show the gas-phase ISM metallicities of SDSS galaxies (obtained from the MPA/JHU catalogue) with stellar masses $> 10^{10}\,\msun$ and with $<10^{7}\,\msun$.  The $>10^{10}\,\msun$ ISM metallicities are the closest match to our old stellar metallicities, which is not surprising given they trace similar environments.

\begin{figure}
    \centering
    \includegraphics[width=\columnwidth]{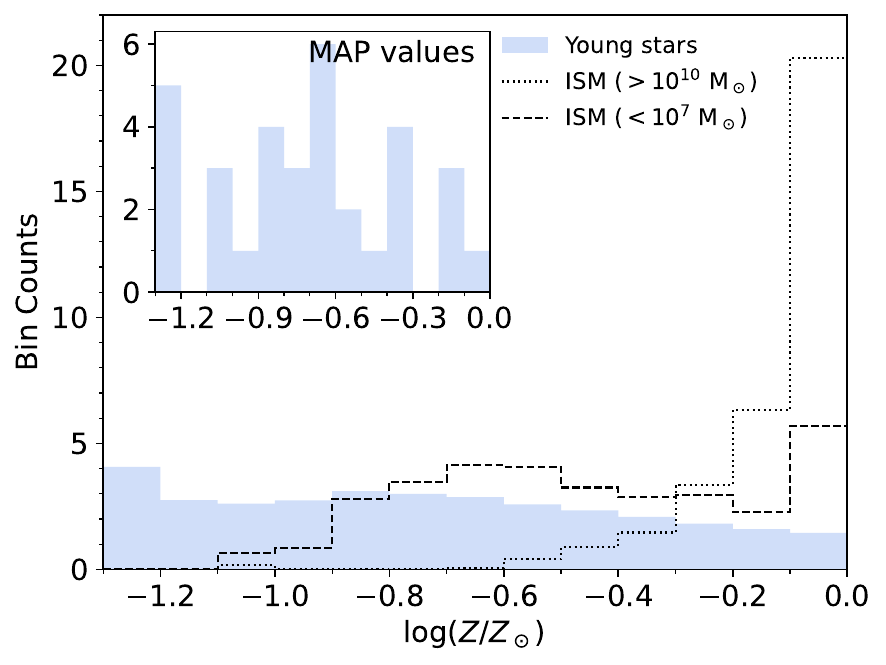}
    \caption{Equivalent to Figure \ref{fig:logzsol_old_posterior}, but for the young stellar population.}
    \label{fig:logzsol_yng_posterior}
\end{figure}

The young stellar population's metallicities are shown in Figure \ref{fig:logzsol_yng_posterior}, overlaid with the same reference distributions as the old population.  We see that the young stars have a much wider distribution than the old stars, covering our full parameter space from $\log(Z_{*,\rm yng}/Z_\odot) = -1.3$ to $0$, though with a slight preference for more metal-poor values.  This width, while it may be indicative of a large intrinsic variation in the young stellar metallicites, is also largely created because our constraints on each individual system's young stellar metallicity are much looser than the old stellar metallicities.  Like with the SFRs, these constraints have a strong dependence on $f_{\rm young}$, with systems below $f_{\rm young}<0.1$ having uncertainties of $\sim 0.3$--$0.4$ dex, and systems above $f_{\rm young}>0.1$ able to get down to $\sim 0.1$ dex. Visually, the distribution more closely matches the $<10^{7}\,\msun$ ISM gas-phase metallicities.  

A KS test between the old and young stellar metallicity distributions reveals a strong $7\sigma$ rejection that they are drawn from the same underlying distribution.  However, unlike the previous KS tests we have presented until this point, for this test we must be a bit more careful in the interpretation of the results, because the two sample distributions we are testing have large, systematic differences in their measurement uncertainties.  

To ensure the KS test result is not an artifact of the larger uncertainties on the young stellar metallicities, we also perform a marginally more sophisticated test by explicitly modeling the underlying sample distributions.  The models consist of $i$ Gaussian functions with weights $w_i$, means $\mu_i$ and intrinsic scatters $\sigma_{i,{\rm intr}}$, which are added in quadrature with the $j$ measurement uncertainties $\sigma_{j,{\rm meas}}$ when calculating the likelihood $\mathcal{L}$ ($\sigma_{ij}^2 = \sigma_{i,{\rm intr}}^2+\sigma_{j,{\rm meas}}^2$):
\begin{equation}
    \mathcal{L} = \prod_{j}^{n_{\rm meas}}\,\Bigg(\sum_{i}^{n_{\rm modes}}\frac{w_i}{\sqrt{2\pi\sigma_{ij}^2}}\exp\bigg[-\frac{1}{2}\frac{(x_j-\mu_i)^2}{\sigma_{ij}^2}\bigg]\Bigg)
\end{equation}
We decide to model the underlying distributions with $n_{\rm modes}=2$ because this minimizes the Bayesian Information Criterion, ${\rm BIC} = (3n_{\rm modes}-1)\ln n_{\rm meas} - \ln \mathcal{L}$ \citep{1978AnSta...6..461S}.  The null hypothesis---that both old and young stellar metallicities can be modeled with the same underlying 2-Gaussian distribution---is then compared to the hypothesis that they must have unique distributions with a likelihood ratio test.  Despite this test's much more careful treatment, we ultimately find that it produces only a marginally less significant result than the KS test, rejecting the null hypothesis at $\sim 6\sigma$ confidence.  Therefore, we can be extremely confident in our conclusion that the old and young stellar metallicities are not drawn from the same distribution.  We further discuss the implications of these results, with the additional context of the CGM metallicities from our \textit{Chandra} analysis, in the discussion ($\S$\ref{sec:discussion_metals}).

\subsection{Dust content} \label{sec:results_dust}

\begin{figure}
    \centering
    \includegraphics[width=\columnwidth]{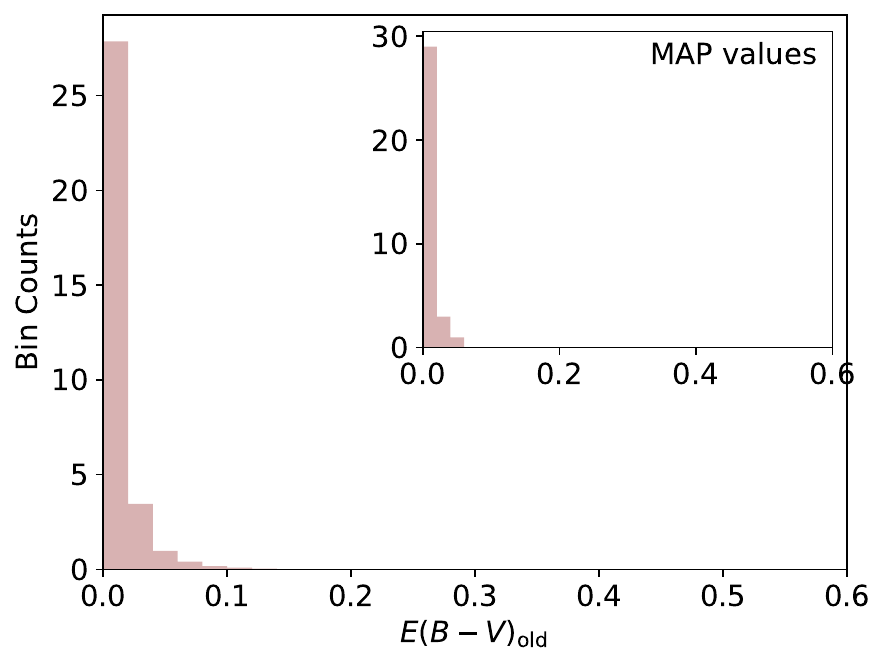}
    \caption{Histograms of the old stellar population's dust reddening, $E(B-V)$.  The main panel shows the cumulative posterior distribution, while the inset panel shows the discretized histogram of maximum a posteriori (MAP) values.}
    \label{fig:dust2_old_posterior}
\end{figure}

\begin{figure}
    \centering
    \includegraphics[width=\columnwidth]{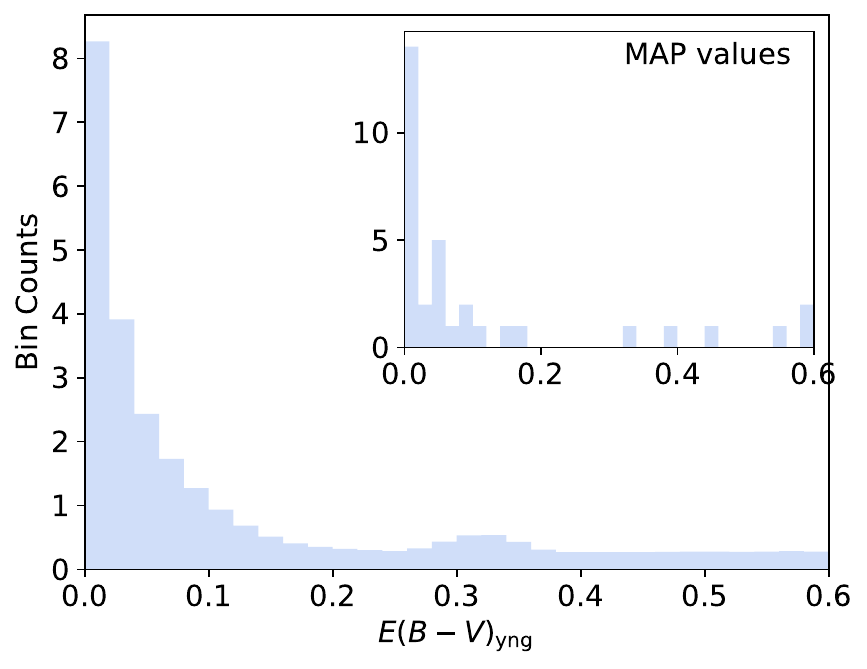}
    \caption{Equivalent to Figure \ref{fig:dust2_old_posterior}, but for the young stellar population.}
    \label{fig:dust2_yng_posterior}
\end{figure}

A histogram of dust reddening, $E(B-V)$, for the old stellar populations is shown in Figure \ref{fig:dust2_old_posterior}.  Internally, our SED fitting routine parametrizes the dust obscuration in terms of the optical depth in the visual band, $\tilde{\tau}_{\rm V}$.  This is related to the visual extinction in magnitudes, $A_{\rm V}$, by
\begin{equation}
    A_{\rm V} = \frac{2.5}{\ln10}\tilde{\tau}_{\rm V} \approx 1.086\tilde{\tau}_{\rm V}~.
\end{equation}
Then, the visual extinction is related to the reddening by $E(B-V)=A_{\rm V}/R_{\rm V}$, where $R_{\rm V} \equiv 4.05$ for the Calzetti extinction curve that we use.  The reddening in the old stellar populations primarily probes the diffuse dust content of our elliptical galaxies, as the old stars themselves are expected to be mostly unobscured and uniformly distributed throughout the galaxies.  The distribution reveals that there is very little diffuse dust, with only a couple of systems falling outside the innermost bin covering $E(B-V) \leqslant 0.02$ mag.  We can turn these measurements into an implied Hydrogen column density using the conversion from \citet{1978ApJ...224..132B}:
\begin{equation}
    N_{\rm H} = (5.8 \times 10^{21}\,{\rm cm^{-2}\,mag^{-1}\,})\,E(B-V)
\end{equation}
This implies that our sample has H columns of $\leqslant 10^{20}$ cm$^{-2}$, which is generally in line with what is seen in the larger elliptical galaxy population \citep[e.g.][]{2012MNRAS.422.1835S}.

\begin{figure}
    \centering
    \includegraphics[width=\columnwidth]{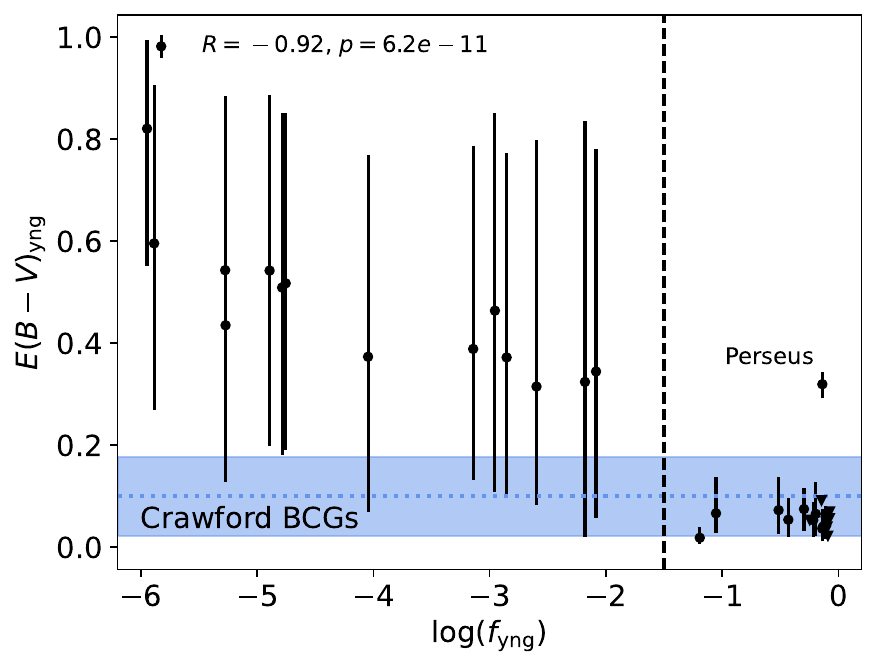}
    \caption{The young stellar population's reddening, $E(B-V)_{\rm yng}$, is plotted against the young fraction $f_{\rm young}$ (eq \eqref{eq:fyoung}).  A vertical dashed line at $\log(f_{\rm young}) = -1.5$ splits the $E(B-V)_{\rm yng}$ values into 2 distinct populations.  The legend gives the Pearson correlation coefficient $R$ and $p$-value, showing a highly significant anticorrelation.  The blue dotted line and blue shaded region show the average and standard deviation of a sample of brightest cluster galaxies (BCGs) from \citet{1999MNRAS.306..857C}.}
    \label{fig:ebv_fyng}
\end{figure}

The young stellar population's obscuration similarly favors smaller values, but with a wider spread (Figure \ref{fig:dust2_yng_posterior}).  Indeed, our ability to constrain the dust in the young stellar populations, much like the previously discussed SFRs and metallicities, is highly dependent on $f_{\rm young}$.  We show this relationship explicitly in Figure \ref{fig:ebv_fyng}, where we plot $E(B-V)_{\rm yng}$ against $f_{\rm young}$.  A cutoff of $\log f_{\rm young} \sim -1.5$ cleanly splits the population into those with good constraints and those without.  The systems without good constraints are uniformly spread from 0--1, indicating that they are simply recovering the uniform prior.  This is observed as a low-amplitude flat component in Figure \ref{fig:dust2_yng_posterior}, superimposed over the systems with better constraints.  As might be expected, a $\chi^2$ test between these systems with $f_{\rm young} < 0.03$ and the flat prior distribution indicates perfect agreement ($0\sigma$).  On the other hand, the systems with $f_{\rm young} > 0.03$ are statistically distinct from the prior distribution at $9\sigma$, though they are only distinct from 0 at $\sim 1.7\sigma$.
% which corresponds to H columns of $10^{20}$--$10^{21}$ cm$^{-2}$. 

Also shown in Figure \ref{fig:ebv_fyng} is a comparison with intrinsic $E(B-V)$ values from a sample of brightest cluster galaxies (BCGs) from \citet{1999MNRAS.306..857C} (hereafter C99).  Care must be taken in this comparison, as C99 use an $R_{\rm V}$ of 3.2, compared to our 4.05.  We correct for this by multiplying their results by $3.2/4.05$.  Additionally, they measure reddenings from the Balmer decrement, which is sensitive to the reddening of the ISM gas, as opposed to the stellar reddening.  We use the standard conversion factor $E(B-V)_{*}=0.44E(B-V)_{\rm gas}$ from \citet{2000ApJ...533..682C}.  The appropriateness of using this particular conversion factor is discussed further in $\S$\ref{sec:discussion_dust}, but we note here that a conversion \textit{is} necessary---we do not expect our young stellar $E(B-V)_{\rm yng}$ to trace 1:1 with $E(B-V)_{\rm gas}$, because the young stellar component may include stars as old as 1 Gyr, much older than the typical birth cloud escape timescale of $\sim 10$ Myr.  The average and standard deviation of their sample are shown in blue.  In general, our results fall below their average (for the systems where it is well measured), but are within 1 standard deviation.  One system (Perseus) is an exception, with a high young fraction and high $E(B-V)_{\rm yng} = 0.32 \pm 0.03$ that is well constrained.  Otherwise, we find an average young stellar reddening of $E(B-V)_{\rm yng} = 0.04 \pm 0.02({\rm std})$.  There is one individual system covered by both C99 and our sample where we both have good constraints, Abell 1795.  For this system, they find a (converted) $E(B-V) = 0.05^{+0.05}_{-0.05}$, which is in good agreement with our result of $E(B-V)_{\rm yng} = 0.06^{+0.06}_{-0.04}$.

\section{Discussion} \label{sec:discussion}

% split into a few topics

\subsection{The UV upturn in massive elliptical galaxies} \label{sec:discussion_uv_upturn}

With this study's focus on the far-UV emission of massive elliptical galaxies, it would be remiss not to include a discussion on the UV upturn phenomenon, despite it not being the main goal of our analysis.  Many of the galaxies in our sample are classic UV upturn examples (i.e. NGC 1399, NGC 1404, NGC 4374, NGC 4552), and 24/30 systems exhibit $FUV-V$ colors $\gtrsim 6.5$ mag, indicative of enhanced UV emission over the expected amount from an old stellar population \citep{2007ApJS..173..607R}.  

As mentioned in $\S$\ref{sec:templates}, there is a general consensus that the origin of this excess UV emission comes from hot horizontal branch stars and their progeny \citep{1993ApJ...419..596D, 1997ApJ...482..685B}.  However, there are still many unknowns when delving further into the details.  In particular, excess UV emission could be produced by extremely metal-poor HB stars or by metal-rich HB stars \citep{2008ASPC..392....3Y}.  The metal-poor scenario involves standard stellar evolution pathways, but relies on uncomfortably large fractions of metal-poor stars that are difficult to reproduce with standard chemical evolution.  On the other hand, the metal-rich scenario involves stellar evolution onto the EHB and AGB-manqu\'e branches \citep{1990ApJ...364...35G}, which could result from large mass loss rates or enhanced helium abundances \citep{1992ApJ...388L..53H, 1997ApJ...486..201Y}.  This picture is further complicated by the effects of $\alpha$-enhancement, which can make metallicity relationships difficult to interpret, and binary evolution, which may be substantially different from single-star evolution.  

We will not attempt to provide an answer to these questions in this work.  The UV upturn strength is handled in our SED modeling procedure with the $f_{\rm EHB}$ parameter, which allows for a contribution from a population of metal-rich EHB stars.  The uncertainty in the decomposition between a young stellar population and EHB stars from the old stellar population is already quite large in most systems with this setup, which tells us that the data we have are not constraining enough to answer these questions.  Adding another model parameter to allow the EHB star's metallicity to vary would likely only further muddle the results without providing any clear answers.  Indeed, the primary observational probe that has been used to investigate this problem in the past is the evolution of $FUV-V$ color with redshift \citep{2007ApJS..173..607R}, but we are not positioned to examine this trend either, as our sample consists exclusively of local systems at $z < 0.1$.

\begin{figure}
    \centering
    \includegraphics[width=\columnwidth]{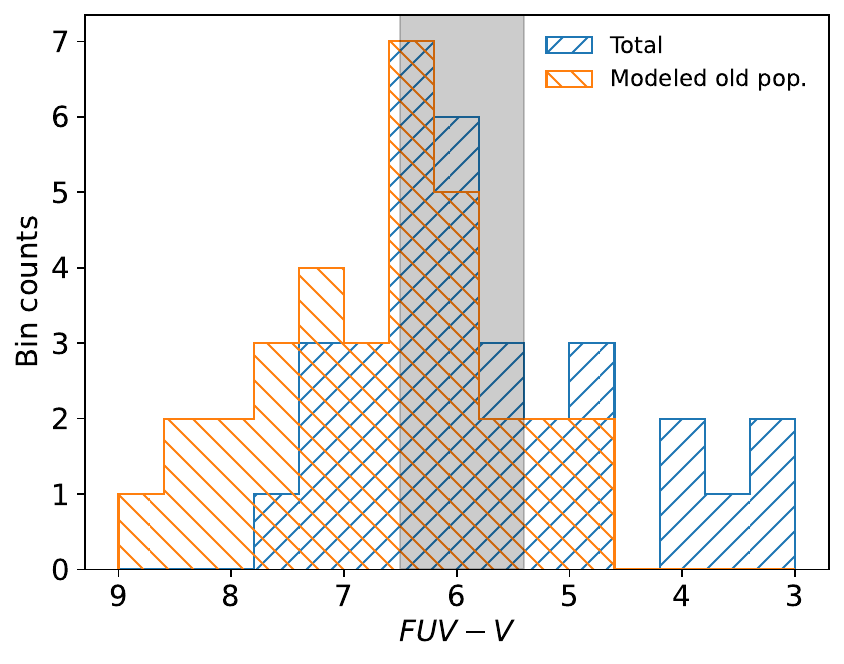}
    \caption{A histogram of $FUV-V$ color for the galaxies in our sample, with a bin size of 0.4 mag.  The shaded gray band shows $5.4 < FUV-V < 6.5$, colors typically associated with UV upturn galaxies.  The observed photometry of our systems in shown in the blue forward-hatched histogram, and the model photometry using only the old stellar population component is shown in the orange backward-hatched histogram.}
    \label{fig:uv_colors}
\end{figure}

An interesting question we \textit{can} answer, while keeping the uncertainty in stellar population decomposition in mind as an important caveat, is how our inferred young stellar population contribution affects the $FUV-V$ color (and therefore the UV upturn strength).  In Figure \ref{fig:uv_colors} we histogram of our sample's $FUV-V$ colors.  Since we do not have a direct analogue of the $V$ band, these colors are extracted from our best-fitting SED models using \textit{GALEX} FUV and Johnson $V$ filter transmission curves.  The total colors (blue histogram) show that most of our sample falls within the gray band covering $5.4 \lesssim FUV-V \lesssim 6.5$, colors typically associated with the UV upturn \citep{2007ApJS..173..607R}, with some outliers at bluer colors due to contamination from ongoing star formation (e.g. Perseus, Abell 1795, Abell 2597, M87).  When colors are instead taken from only the old stellar population component (orange histogram), the colors generally shift to the left (redder). This makes sense, as they lose a strong blue component from the young stars. As a consequence, we see many of the strongly star-forming systems get pushed into the gray UV upturn band.  In other words, we find that a UV upturn is not mutually exclusive with ongoing active star formation, as these systems exhibit both features simultaneously. 

Interestingly, we also observe a large handful of systems get pushed \textit{out} of the UV upturn band, into the quiescent region.  This means that our SED modeling preferred to fit the excess UV emission in these galaxies with low levels of ongoing star formation, rather than a UV upturn component in the old population.  If this trend is physical, and not just an artifact of our modeling uncertainties, it points to an important conclusion: $FUV-V$ color diagnostics are \textit{not} a clean separator of UV upturn galaxies from low-level star-forming galaxies, and more careful modeling must be done to adequately disentangle the two.

\begin{figure}
    \centering
    \includegraphics[width=\columnwidth]{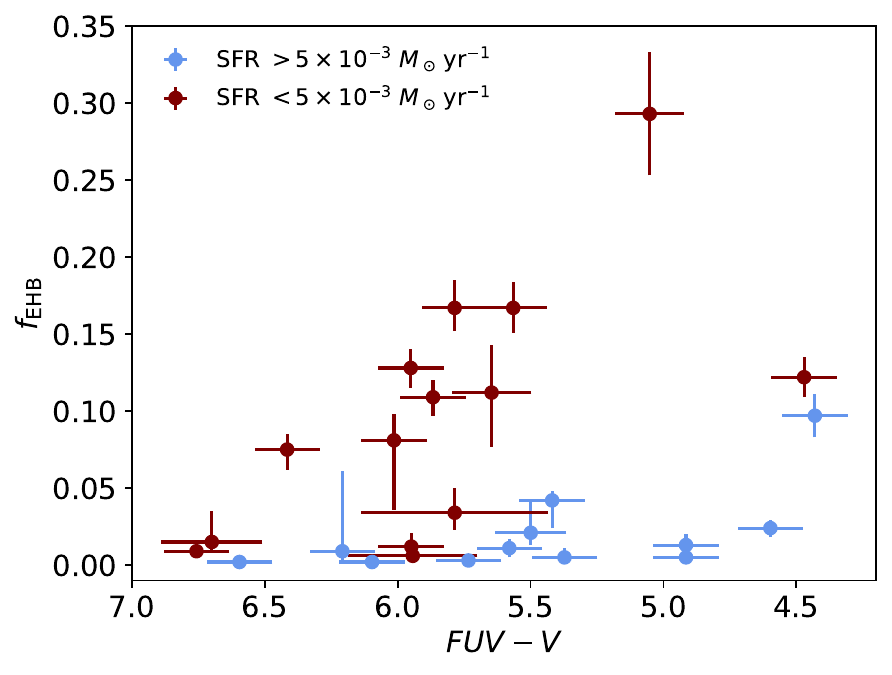}
    \caption{EHB fraction ($f_{\rm EHB}$ as a function of observed UV upturn strength ($FUV-V$ color).  Blue points show systems with SFRs $> 5 \times 10^{-3}$ \msunyr\, and red points show systems with SFRs $< 5 \times 10^{-3}$ \msunyr.}
    \label{fig:fehb_fuv_v}
\end{figure}

As a result of this phenomenon, we also observe a bimodality in our recovered $f_{\rm EHB}$ parameter as a function of UV upturn strength (parametrized by the total $FUV-V$ color), which we show in Figure \ref{fig:fehb_fuv_v}.  The systems with substantial EHB contributions (shown in red) generally follow a power law, with $f_{\rm EHB}$ increasing by $\sim 0.3$ over 2 mag in $FUV-V$.  Whereas the systems modeled with mostly young populations (shown in blue) have $f_{\rm EHB} \sim 0$, independent of $FUV-V$.  This implies a large intrinsic variation in the relative quantity of EHB stars in massive elliptical galaxies of anywhere from $\sim 0$--$30$\%, which presents another interesting puzzle for stellar evolution.  This variation may be explained by differences in metallicity \citep{1998PhDT.........5D}.

\subsection{The origins of the youngest stars in massive elliptical galaxies} \label{sec:discussion_metals}

We now aim to understand the origin of the observed ``residual'' levels of star formation in these otherwise quiescent massive elliptical galaxies (from $\log({\rm SFR}/\msunyr) \sim -3$ to $1$).  The gas that these stars form from could have a few possible origins:
\begin{enumerate}
    \item Recycled gas from old stars, blown out from stellar feedback within the galaxy
    \item Gas that has cooled out of the hot phase of the CGM and been deposited into the galaxy
    \item Stripped cool gas from nearby or merging dwarf galaxies
\end{enumerate}
These scenarios can be distinguished by the metallicity imprint that they leave on the young stellar population.  In the rest of this subsection, we will go through each of these scenarios one-by-one and discuss their feasibility based on the data.

It is important to preface this discussion with a clarification that our metallicity measurements of the old stars, young stars, and CGM gas are measured over different timescales, so care must be taken when comparing them to each other \citep[see e.g.][]{2022MNRAS.510..320F}.  The old stellar metallicity is a luminosity-weighted average of the metallicity within all stars formed $>$ 1 Gyr ago, while the young stellar metallicity is a luminosity-weighted average of the stars formed $<$ 1 Gyr ago (with the 1 Gyr cutoff being a consequence of how we define our model's SFH).  In contrast, the CGM gas metallicity measured with \textit{Chandra} probes the gas' metal enrichment at the \textit{current epoch} (in the galaxy's rest frame).

In the simplest scenario (1), gas gets recycled from the old stellar population, without any external gas inflows or outflows.  This is also known as the ``closed box'' scenario, within which the metal enrichment history should monotonically increase over time due to the byproducts of stellar evolution.  Since the old stellar metallicity is an average over a much longer timescale than the young stellar metallicity, we should expect the young stellar metallicity to be higher than the old stellar metallicity in this scenario, and lower than the CGM gas metallicity: $Z_{\rm *,old} < Z_{\rm *,yng} < Z_{\rm CGM}$.  We can therefore immediately reject this scenario as the primary formation channel in our sample---not only do we observe almost exclusively the opposite trend, with $Z_{\rm *,yng} \lesssim Z_{\rm *,old}$ (see, e.g., Figures \ref{fig:logzsol_old_posterior} and \ref{fig:logzsol_yng_posterior}), but this trend is extremely significant at the $6\sigma$ level, as confirmed by our rigorous statistical testing in $\S$\ref{sec:results_logz}.

In scenarios (2) and (3), inflows of gas are introduced from different sources that can cause the young stellar metallicity to be lower than the old stellar metallicity.  In scenario (2), the origin of this gas is the CGM of the galaxy itself (i.e. a cooling flow), which implies that $Z_{\rm CGM}$ should be correlated with $Z_{\rm *,yng}$ in such a way that $Z_{\rm *,yng} \sim Z_{\rm CGM}$ (if one considers the CGM a large enough reservoir here that any metal enrichment from stellar feedback is negligible).  

\begin{figure}
    \centering
    \includegraphics[width=\columnwidth]{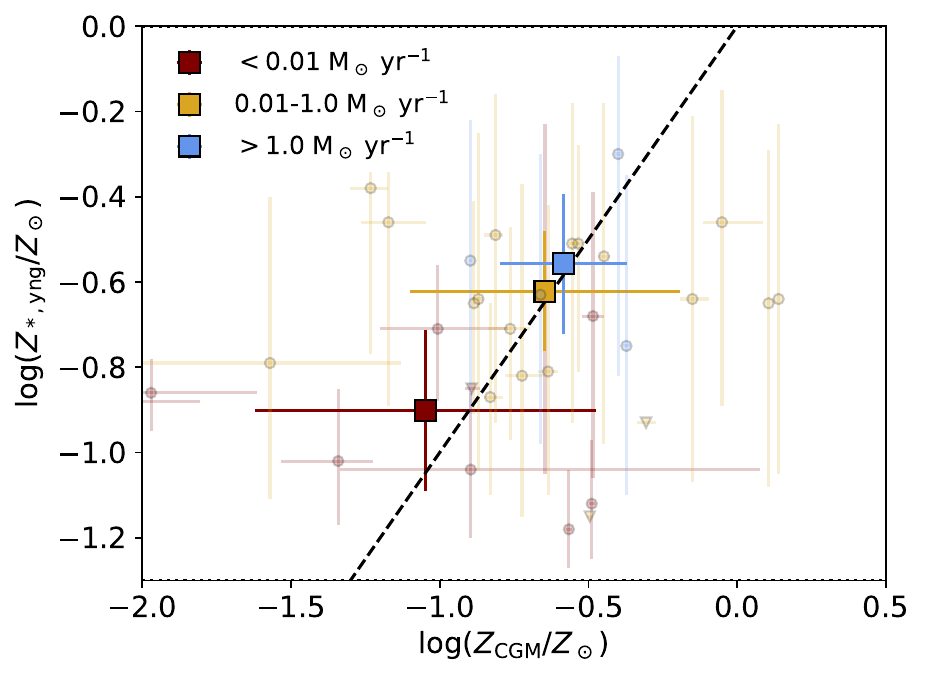}
    \caption{A scatterplot showing the young stellar metallicity $Z_{\rm *,yng}$ against the CGM metallicity $Z_{\rm CGM}$. The data is grouped into 2-dex-wide bins in SFR, shown by the different colors given in the legend.  Individual systems are shown in the background with a low opacity, while binned data points with the mean and scatter within each bin are shown in full opacity in large square points.  The dashed line shows where the two metallicities are equal.  It is observed that the binned data converges towards the line of equality as the constraints on the data improve (with higher SFR).}
    \label{fig:zdiff}
\end{figure}

In Figure \ref{fig:zdiff}, we plot $Z_{\rm *,yng}$ against $Z_{\rm CGM}$, where $Z_{\rm CGM}$ is taken from the cooler temperature component of our 2-temperature X-ray spectral models.  There is a large scatter in $Z_{\rm *,yng}$, driven primarily by our statistical and systematic uncertainties, but a degree of physical scatter is also expected if star formation is driven by multiple scenarios, rather than a single one. This effect will be the greatest for the systems with the smallest SFRs, where our constraints are the weakest.  We therefore bin the data into 3 SFR bins, which roughly equates to binning by the signal-to-noise ratio, and observe a clear trend where the higher-quality data (higher SFR) converges toward the line of equality, shown by the dashed line.  This provides evidence that these two metallicities are similar to each other, both having a sub-solar value of $-0.5$ to $-0.6$ on average.  In other words, this indicates that the coolest CGM gas is closely linked with the youngest stars.

On the other hand, scenario (3) proposes that the origin of the low-metallicity gas instead comes from accreted low-mass systems.  It is difficult to test this scenario, because we do not have a direct observational probe of the metallicities of individual systems which may or may not have been accreted in the past history of these giant elliptical galaxies.  However, based on knowledge of the typical metallicities seen in such systems, we can infer that this scenario is also plausible based on our results. 

\begin{figure}
    \centering
    \includegraphics[width=\columnwidth]{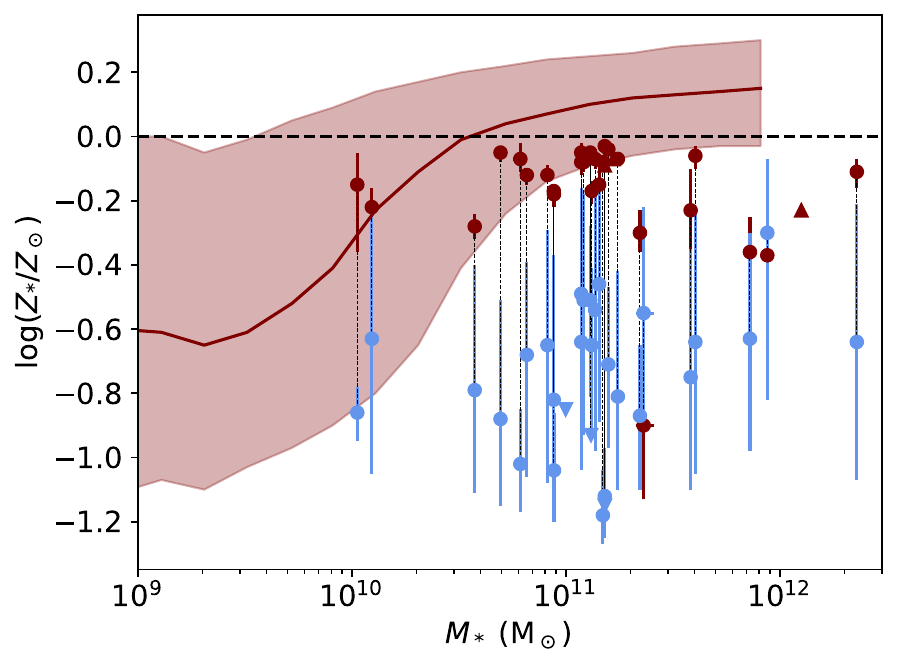}
    \caption{The stellar mass--stellar metallicity ($M_*-Z_*$) relation from \citet{2005MNRAS.362...41G} is shown with the red line, with the $1\sigma$ scatter shown with the red shaded region.  The metallicities of our old stellar population, $Z_{\rm *,old}$, are shown in the red circles, and the metallicities of our young stellar population, $Z_{\rm *,yng}$, are shown in the blue circles.  The values of $Z_{\rm *,old}$ and $Z_{\rm *,yng}$ for the same galaxy are connected by thin black dashed lines.  The thicker horizontal dashed black line notes the imposed upper limit on our metallicities, due to the limited grid space coverage of our stellar templates, which likely biases our $Z_{\rm *,old}$ results low.}
    \label{fig:mz_relation}
\end{figure}

Referring back to Figure \ref{fig:logzsol_yng_posterior}, the histogram of ISM metallicities from SDSS galaxies with $M_* \leqslant 10^7\,\msun$ is a relatively close match to the observed $Z_{\rm *,yng}$ distribution.  Furthermore, $Z_{\rm *,yng}$ tends to be very consistent with the abundance pattern that would be expected from $10^{10}\,\msun$ and below galaxies from the mass-metallicity relation, while the old stellar metallicities are much more enriched, as is typical of $\gtrsim 10^{10}\,\msun$ galaxies (excluding the aforementioned outlier systems).  We show the mass-metallicity relation in Figure \ref{fig:mz_relation}, highlighting the difference between the old and young populations, and noting the upper limit of $\log (Z_{\rm *}/Z_\odot) \leqslant 0$ with the dashed black line, which likely biases some of our old stellar metallicities to be lower than they truly are. This shows that the young stellar populations are consistent with being fueled by stripped gas from $<10^{10}\,\msun$ satellites.

Although we are unable to distinguish the source (CGM deposition or external accretion from merging systems), it nevertheless seems clear that the recently formed young stars are fueled by a reservoir of (relatively) metal-poor gas that is distinct from the source of the old stellar population.  This result is robust, even in light of the potential systematic effects in our measurements, and it is in agreement with previous works \citep[i.e.][]{2019MNRAS.489L.108D}, which have found that roughly $\sim 37\%$ of elliptical galaxies have cool gas metallicities distinct from stellar metallicities, which they interpret as being fueled by mergers and accretion.  Notably, 37\% is close to the cool-core fraction \citep{2026ApJ..1000..252W}.  And, if the CGM is indeed the source of the stellar fuel, it is fully consistent with a reduced cooling flow scenario.

\subsection{The geometry and density of dust in massive elliptical galaxies} \label{sec:discussion_dust}

% - pivot this section to talk more about the interpretation of the difference between our E(B-V) values and Crawford's

It is well established that the ISM of galaxies has a ``clumpy'' nature, with star formation proceeding primarily in the coldest, densest regions of molecular gas \citep{2010ApJ...724..687L}.  These regions also tend to be the most dusty \citep[for a recent review, see][]{2025arXiv250401410S}.  In this section, we aim to understand the implications of our $E(B-V)$ values, regarding both the (relative) geometric distribution and the (absolute) density of the dust.

\begin{figure}
    \centering
    \includegraphics[width=\columnwidth]{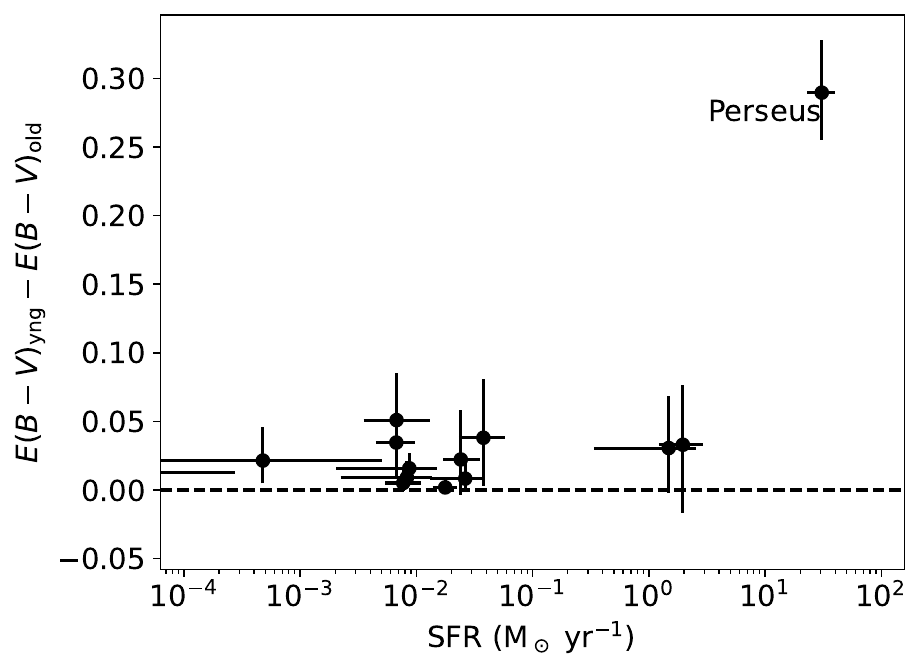}
    \caption{The difference in reddening between the young and old stellar populations, $E(B-V)_{\rm yng}-E(B-V)_{\rm old}$, is shown as a function of the SFR.  Only systems with young fractions $f_{\rm young} > 0.03$ are shown.  A dashed line marks 0, where the two reddenings are equal.  Perseus is labeled as a significant outlier to the rest of the sample.}
    \label{fig:ebv_diff}
\end{figure}

Geometrically, the clumpy nature of the ISM would predict that the young stellar population's reddening should be larger than the old population's. In Figure \ref{fig:ebv_diff}, we show the difference in our young and old stellar population's reddening, $\Delta E(B-V) \equiv E(B-V)_{\rm yng}-E(B-V)_{\rm old}$.  Only systems with young fractions $f_{\rm young} > 0.03$ are shown, as the rest of the systems have poorly constrained $E(B-V)_{\rm yng}$.  We observe that all systems recover $\Delta E(B-V)>0$, in support of a clumpy ISM, with an average (excluding Perseus as an outlier) of $\Delta E(B-V) = 0.007 \pm 0.003$. Of the 4 systems in our sample with multiple \textit{FUSE} pointings, 2 systems (Abell 1795 and NGC 4636) have physically unique, non-overlapping pointings from which we can study spatial variations in the dust content.  In both of these systems, we find little evidence for variation in the dust reddening between pointings.  The large $30'' \times 30''$ \textit{FUSE} aperture size is too large to resolve individual dust structures in any of our galaxies, so large variations between pointings is not necessarily expected.

\begin{figure}
    \centering
    \includegraphics[width=\columnwidth]{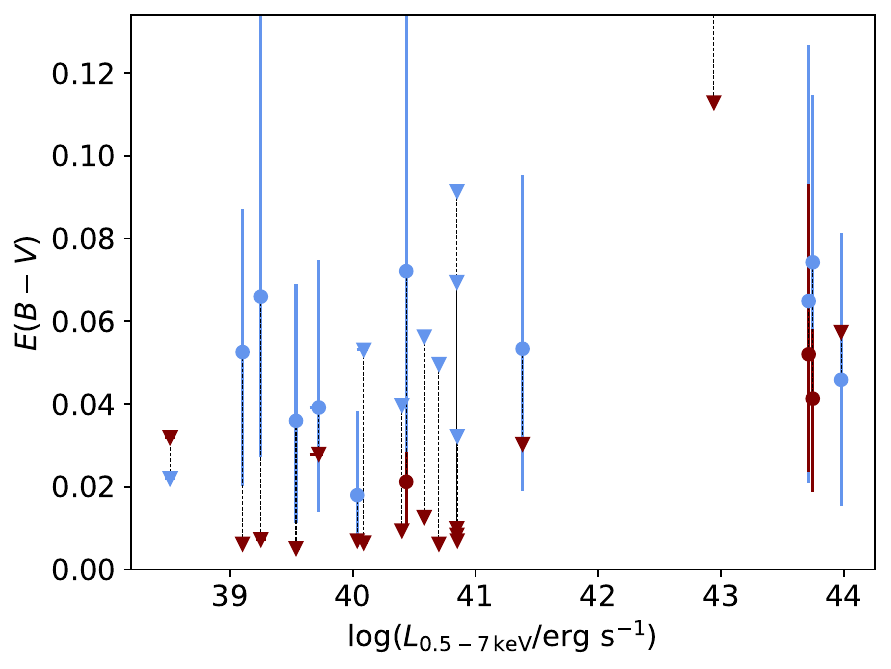}
    \caption{The reddening $E(B-V)$ is shown as a function of total X-ray luminosity integrated from 0.5--7 keV.  Red points show the old population, and blue points show the young population.  Only systems with a young fraction $f_{\rm young} > 0.03$ are shown.  Thin dashed black lines connect the points for the same galaxy.  No significant trends are observed.}
    \label{fig:LX_ebv}
\end{figure}

We do not observe any significant trends in $E(B-V)$ with stellar mass or X-ray luminosity (Figure \ref{fig:LX_ebv}).  It is undeniable that environment plays a role in dust formation, because it is well known that spiral galaxies show categorically different dust distributions from elliptical galaxies.  However, our results indicate that this environmental influence may ``saturate'' above a threshold of $\sim 10^{10}\,\msun$, when the most important factors that determine the accumulation of dust are no longer dictated by the environment.  It is likely that microphysics plays a significant role in determining the formation of cold clumps of gas and dust, which allow the more brittle grains to be effectively self-shielded from the harsh external radiation field which would otherwise easily sputter them within $\lesssim 10^7$ yr.  The total amount of dust in these galaxies, then, would be primarily determined by how easy it is to form and sustain these cold clumps, which is influenced by the thermodynamic and magnetic properties of the halo.

Regarding the absolute dust content of these systems, we return to our comparison with the sample of BCGs from C99, presented in $\S$\ref{sec:results_dust}.  While, as established, the young stellar $E(B-V)$ values of our sample are generally within 1 standard deviation of C99's sample, the means of the two samples are significantly different:
we find average young stellar reddening of $0.040 \pm 0.010$, which is smaller than the average in the C99 sample of $0.099 \pm 0.005$ by $\sim6\sigma$.  There are a few possible explanations for this discrepancy.  First, the C99 average may be biased high due to the preferential inclusion of systems for which they were able to measure a detectable $E(B-V)$. Our sample is affected by this bias as well, but our ability to measure $E(B-V)$ is primarily driven by the brightness of the young stellar population ($f_{\rm young}$) as opposed to the magnitude of $E(B-V)$, so our average should not be driven in one direction or the other.  Second, assuming the difference is physical, our average could be driven lower due to differences in our sample populations---our sample includes lower-mass group and isolated elliptical halos, whereas C99 studies a pure sample of X-ray-bright cluster BCGs.  If this is the case, it would indicate that more massive cluster halos tend to host more dust than group and elliptical halos, despite their hotter halo temperatures.  This challenges our earlier finding that $E(B-V)_{\rm yng}$ does not significantly evolve with stellar mass or X-ray luminosity, so, if such a trend does exist, it must be below our detection threshold.  Another possibility is that the standard conversion of $E(B-V)_*=0.44E(B-V)_{\rm gas}$, which has been calibrated on actively star-forming and starburst galaxies, is not appropriate to use for massive elliptical galaxies, which may have substantially different dust properties.  If we take C99's $E(B-V)_{\rm gas}$ values, along with our stellar $E(B-V)_{\rm yng}$ values, at face value, we would instead derive a much lower conversion factor of
\begin{align}
    E(B-V)_{\rm yng} &= (0.17 \pm 0.05)E(B-V)_{\rm gas} \\
    E(B-V)_{\rm old} &\leqslant 0.05E(B-V)_{\rm gas}
\end{align}
which indicates that the dust distribution is much more concentrated around \ion{H}{2} regions than in typical star-forming galaxies.

% \subsection{A new perspective on cooling flows in massive elliptical galaxies}

\section{Conclusions} \label{sec:conclusion}

We have laid the groundwork for an in-depth homogeneous reanalysis of the \textit{FUSE} archive of 29 massive elliptical galaxies, which we plan to fully cover over a series of publications.  In this first paper of the series, we have re-reduced the \textit{FUSE} spectra with the most modern version of the CalFUSE reduction software, following the most up-to-date recommendations by the developers to maximize the signal-to-noise of the output spectrum and robustly estimate the background.  In tandem, we have obtained aperture-matched and absolute-flux-calibrated photometry from the deepest available \textit{GALEX}, SDSS, DES/DECaLS, 2MASS, and \textit{WISE} data, as well as the deepest aperture-matched X-ray spectra from \textit{Chandra}.  We have also gathered the most up-to-date canonical information on these systems' redshifts and physical distances from the NED archive, allowing for the calculation of the most precise luminosities (and related quantities, i.e. stellar mass). 

We have additionally compiled a complete modern set of stellar template libraries across $T_{\rm eff}$, $\log g$, and $\log Z/Z_\odot$, homogenized to a standard solar metallicity from \citet{2009ARAnA..47..481A}, and at extremely high resolution in the FUV ($\Delta\lambda \lesssim 0.2\,\angstrom$), while covering a full spectral range from $90$ \angstrom--$1000$ \um~ at lower resolution.  These have come with the additional gathering, inclusion, and physically motivated treatment of extended isochrones / evolutionary tracks for so-called ``extreme'' horizontal branch (EHB) stars, which make up a substantial portion of the observed FUV emission in elliptical galaxies, contributing to the ``UV upturn'' phenomenon.  These come in tandem with updates to the \textsc{FSPS} source code to accommodate the new libraries, which are made available through forks of \textsc{FSPS} and \textsc{Python FSPS} on GitHub.

In this first paper, we have used these data and templates to model the stellar populations of these 29 massive elliptical galaxies, leveraging our multiwavelength data in the FUV--MIR to simultaneously constrain the stellar masses, star formation histories, stellar metallicities, dust obscuration, and EHB fraction.  This analysis has produced a number of important results:
\begin{itemize}
    \item We recover statistically significant SFRs down to $\sim 10^{-3}\,\msunyr$, driving the observational floor deeper than in previous studies, and find a significant deviation from canonical calibrated scaling relations with 2-band \textit{GALEX} and \textit{WISE} colors in a large fraction of systems due to the UV upturn.
    \item The stellar age and decay time posteriors suggest that, after their initial episode of rapid star formation, most massive ellipticals continue to experience low levels of $\sim$constant star formation over much of their histories, superimposed with short bursts of activity.  This supports the picture of bursty feeding and feedback which has been established in previous works \citep{2012ApJ...746...94G, 2013MNRAS.432.3401G}.
    \item The metallicities of the old and young stellar populations strongly suggest that they cannot have formed from the same gas, effectively ruling out recycled gas as a significant contributor to recent star formation.
    \item The $E(B-V)$ seen in the old and young stellar populations reinforce the well established picture that the ISM is clumpy, while a comparison with \citet{1999MNRAS.306..857C} hints at a deviation from the starburst-like stellar-to-gas $E(B-V)$ ratio from 0.44 to 0.17 and below.
    \item We find that $FUV-V$ colors are a potentially problematic diagnostic to use for separating quiescent galaxies with strong UV upturns from galaxies with residual levels of ongoing star formation, as both types of systems exhibit indistinguishable colors.
\end{itemize}

In parallel, we have used the X-ray data from \textit{Chandra} to infer the metallicity properties of the diffuse gas in the circumgalactic  and intracluser media (CGM/ICM) surrounding the galaxies, and analyzed its correlation with the young stellar metallicities.  We find evidence that these metallicities are correlated, with the correlation being tighter for the galaxies with the highest star formation rates, indicating that these stars may be formed directly out of the cooling CGM.  At the same time, the young stellar metallicities form a distribution which is appropriate for low metallicity dwarf galaxies, suggesting that the source of their fuel may instead be small accreted/merging systems.  

Unfortunately, we cannot further distinguish the origin of the young stars between these two scenarios, which will require a larger sample of better constrained young stellar metallicities to confirm.  Right now, the selection of instruments with FUV capabilities is limited, but using data from \textit{HST}'s \textit{Cosmic Origins Spectrograph}, or upcoming and planned instruments like \textit{Aspera} and \textit{UVEX}, will be invaluable for narrowing down the possibilities and further investigating residual cooling flows.

Finally, the stellar continuum models we have obtained from this work are of high quality ($\chi^2/{\rm dof} \approx 1$), and will be useful in the next paper(s) in this series for modeling the \ion{O}{6} $\lambda\lambda$1032,1038\angstrom~ emission, from which we hope to constrain the amount of gas cooling through $\sim 10^{5.5}$ K and being deposited onto the central galaxies ($\dot{M}_{\textsc{Ovi}}$).  In context with the SFRs from this paper, and newly calculated X-ray cooling rates ($\dot{M}_c$) from the \textit{Chandra} data, we will provide a new perspective on the cooling flow problem across the cluster, group, and elliptical scales.

High-level science products, including the coadded, reduced, and binned \textit{FUSE} spectra, aperture-matched \textit{GALEX}, SDSS, DES/DECaLS, 2MASS, and \textit{WISE} photometry, aperture-matched \textit{Chandra} spectra, and the full table of our SED and X-ray spectroscopy fitting results (Table \ref{tab:big_table}), are made publicly available for download on Zenodo (DOI: \href{https://www.doi.org/10.5281/zenodo.19120699}{10.5281/zenodo.19120699}) to provide the community with an easily accessible tool for further analysis and study.

\begin{deluxetable*}{lLLLLLLLDL}
\tabletypesize{\footnotesize}
\tablecaption{SED fitting results}
\label{tab:big_table}
\tablehead{
    \colhead{Target Name} & \colhead{$t_{\rm age,old}$} & \colhead{$\log(Z_{*,\rm old}/Z_\odot)$} & \colhead{$t_{\rm age,yng}$} & \colhead{$\log(Z_{*,\rm yng}/Z_\odot)$} & \colhead{$f_{\rm EHB}$} & \colhead{$f_{\rm AGN}$} & \colhead{$\sigma_*$} & \twocolhead{SFR$_{10\,{\rm Myr}}$\tablenotemark{a}} & \colhead{...} \\
    \colhead{} & \colhead{Gyr} & \colhead{} & \colhead{Gyr} & \colhead{} & \colhead{} & \colhead{$L_{\rm AGN}/L_*$} & \colhead{\kms} & \twocolhead{\msunyr} & \colhead{...}
}
\decimals
\startdata
Perseus (Abell 426) & 12.39 & -0.90 & 0.222 & -0.55 & 0.347 & 0.042 & 560 & 30.7000 & ... \\
Fornax (Abell S0373) & 13.46 & -0.01 & 0.097 & -1.20 & 0.024 & 0.001 & 280 & 0.0177 & ... \\
Virgo (Center 1) & 13.67 & -0.05 & 0.524 & -0.51 & 0.097 & 0.004 & 590 & 0.0377 & ... \\
Virgo (Center 2) & 13.71 & -0.07 & 0.603 & -0.54 & 0.122 & 0.002 & 660 & 0.0000 & ... \\
Abell 1795 (Center) & 12.69 & -0.37 & 0.544 & -0.30 & 0.357 & 0.003 & 540 & 1.9600 & ... \\
Abell 1795 (Filament) & 11.60 & -0.23 & 0.414 & -0.75 & 0.235 & 0.006 & 370 & 0.0000 & ... \\
Abell 2029 & 11.18 & -0.11 & 0.584 & -0.64 & 0.167 & 0.001 & 330 & 0.0019 & ... \\
Abell 2597 & 11.81 & -0.36 & 0.403 & -0.63 & 0.254 & 0.006 & 520 & 1.4800 & ... \\
Abell 3112 & 10.87 & -0.04 & 0.511 & -0.46 & 0.293 & 0.001 & 320 & 0.0026 & ... \\
AWM 7 & 13.29 & -0.06 & 0.558 & -0.64 & 0.034 & 0.001 & 470 & 0.0038 & ... \\
% Virgo P1/G1 = Center
% Virgo P2/G2 = Jet
% A1795 P1/G1 = Filament
% A1795 P2/G2 = Center
% NGC 4636 P1/G1 = Center
% NGC 4636 P2/G2 = Filament
% NGC 4649 P1/G1 = Center, rot 1 | (they are the same centroid but just have different PAs)
% NGC 4649 P2/G2 = Center, rot 2 |
... & ... & ... & ... & ... & ... & ... & ... & ... & ... \\
\hline
\hline
\enddata
\tablenotetext{}{Only a small subset of columns and rows are shown here.  The full table is available in machine-readable format. Reported values are the 50$^{\rm th}$ percentile of the posterior distribution. The full machine-readable table also includes the 0.3$^{\rm rd}$, 15.9$^{\rm th}$, 84.1$^{\rm st}$, and 99.7$^{\rm th}$ percentiles.}
\tablenotetext{a}{All stellar masses and SFRs derived from our SED models assume a Salpeter IMF.  To convert stellar masses to those that would be measured with a Chabrier or Kroupa IMF, multiply by 0.61 or 0.66, respectively.  To convert SFRs, multiply by 0.63 or 0.67, respectively \citep{2014ARAA..52..415M}.}
\end{deluxetable*}

\begin{acknowledgements}

%%% NSF GRFP %%%

MR acknowledges support from the National Science Foundation Graduate Research Fellowship under Grant No. 2141064.  

%%% NASA ADAP %%%

MR and MM received funding for this project from the NASA Astrophysics Data Analysis Program under grant 80NSSC25K7560.

%%% FUSE and GALEX (legacy text at: https://archive.stsci.edu/publishing/mission-acknowledgements#section-c3c6c066-474e-4a5c-96ca-5fd3fe025b6b) %%%

This research is based on observations made with the \textit{FUSE} and \textit{GALEX} missions, obtained from the MAST data archive at the Space Telescope Science Institute, which is operated by the Association of Universities for Research in Astronomy, Inc., under NASA contract NAS 5–26555.

%%% SDSS %%%

Funding for the Sloan Digital Sky Survey IV has been provided by the Alfred P. Sloan Foundation, the U.S. Department of Energy Office of Science, and the Participating Institutions. SDSS acknowledges support and resources from the Center for High-Performance Computing at the University of Utah. The SDSS web site is www.sdss4.org.

SDSS is managed by the Astrophysical Research Consortium for the Participating Institutions of the SDSS Collaboration including the Brazilian Participation Group, the Carnegie Institution for Science, Carnegie Mellon University, Center for Astrophysics | Harvard \& Smithsonian (CfA), the Chilean Participation Group, the French Participation Group, Instituto de Astrofísica de Canarias, The Johns Hopkins University, Kavli Institute for the Physics and Mathematics of the Universe (IPMU) / University of Tokyo, the Korean Participation Group, Lawrence Berkeley National Laboratory, Leibniz Institut für Astrophysik Potsdam (AIP), Max-Planck-Institut für Astronomie (MPIA Heidelberg), Max-Planck-Institut für Astrophysik (MPA Garching), Max-Planck-Institut für Extraterrestrische Physik (MPE), National Astronomical Observatories of China, New Mexico State University, New York University, University of Notre Dame, Observatório Nacional / MCTI, The Ohio State University, Pennsylvania State University, Shanghai Astronomical Observatory, United Kingdom Participation Group, Universidad Nacional Autónoma de México, University of Arizona, University of Colorado Boulder, University of Oxford, University of Portsmouth, University of Utah, University of Virginia, University of Washington, University of Wisconsin, Vanderbilt University, and Yale University.

%%% DES & DECaLS %%%

The Legacy Surveys consist of three individual and complementary projects: the Dark Energy Camera Legacy Survey (DECaLS; Proposal ID \#2014B-0404; PIs: David Schlegel and Arjun Dey), the Beijing-Arizona Sky Survey (BASS; NOAO Prop. ID \#2015A-0801; PIs: Zhou Xu and Xiaohui Fan), and the Mayall z-band Legacy Survey (MzLS; Prop. ID 
\#2016A-0453; PI: Arjun Dey). DECaLS, BASS and MzLS together include data obtained, respectively, at the Blanco telescope, Cerro Tololo Inter-American Observatory, NSF’s NOIRLab; the Bok telescope, Steward Observatory, University of Arizona; and the Mayall telescope, Kitt Peak National Observatory, NOIRLab. Pipeline processing and analyses of the data were supported by NOIRLab and the Lawrence Berkeley National Laboratory (LBNL). The Legacy Surveys project is honored to be permitted to conduct astronomical research on Iolkam Du’ag (Kitt Peak), a mountain with particular significance to the Tohono O’odham Nation.

NOIRLab is operated by the Association of Universities for Research in Astronomy (AURA) under a cooperative agreement with the National Science Foundation. LBNL is managed by the Regents of the University of California under contract to the U.S. Department of Energy.

This project used data obtained with the Dark Energy Camera (DECam), which was constructed by the Dark Energy Survey (DES) collaboration. Funding for the DES Projects has been provided by the U.S. Department of Energy, the U.S. National Science Foundation, the Ministry of Science and Education of Spain, the Science and Technology Facilities Council of the United Kingdom, the Higher Education Funding Council for England, the National Center for Supercomputing Applications at the University of Illinois at Urbana-Champaign, the Kavli Institute of Cosmological Physics at the University of Chicago, Center for Cosmology and Astro-Particle Physics at the Ohio State University, the Mitchell Institute for Fundamental Physics and Astronomy at Texas A\&M University, Financiadora de Estudos e Projetos, Fundacao Carlos Chagas Filho de Amparo, Financiadora de Estudos e Projetos, Fundacao Carlos Chagas Filho de Amparo a Pesquisa do Estado do Rio de Janeiro, Conselho Nacional de Desenvolvimento Cientifico e Tecnologico and the Ministerio da Ciencia, Tecnologia e Inovacao, the Deutsche Forschungsgemeinschaft and the Collaborating Institutions in the Dark Energy Survey. The Collaborating Institutions are Argonne National Laboratory, the University of California at Santa Cruz, the University of Cambridge, Centro de Investigaciones Energeticas, Medioambientales y Tecnologicas-Madrid, the University of Chicago, University College London, the DES-Brazil Consortium, the University of Edinburgh, the Eidgenossische Technische Hochschule (ETH) Zurich, Fermi National Accelerator Laboratory, the University of Illinois at Urbana-Champaign, the Institut de Ciencies de l’Espai (IEEC/CSIC), the Institut de Fisica d’Altes Energies, Lawrence Berkeley National Laboratory, the Ludwig Maximilians Universitat Munchen and the associated Excellence Cluster Universe, the University of Michigan, NSF’s NOIRLab, the University of Nottingham, the Ohio State University, the University of Pennsylvania, the University of Portsmouth, SLAC National Accelerator Laboratory, Stanford University, the University of Sussex, and Texas A\&M University.

BASS is a key project of the Telescope Access Program (TAP), which has been funded by the National Astronomical Observatories of China, the Chinese Academy of Sciences (the Strategic Priority Research Program “The Emergence of Cosmological Structures” Grant \# XDB09000000), and the Special Fund for Astronomy from the Ministry of Finance. The BASS is also supported by the External Cooperation Program of Chinese Academy of Sciences (Grant \# 114A11KYSB20160057), and Chinese National Natural Science Foundation (Grant \# 12120101003, \# 11433005).

% The Legacy Survey team makes use of data products from the Near-Earth Object Wide-field Infrared Survey Explorer (NEOWISE), which is a project of the Jet Propulsion Laboratory/California Institute of Technology. NEOWISE is funded by the National Aeronautics and Space Administration.

The Legacy Surveys imaging of the DESI footprint is supported by the Director, Office of Science, Office of High Energy Physics of the U.S. Department of Energy under Contract No. DE-AC02-05CH1123, by the National Energy Research Scientific Computing Center, a DOE Office of Science User Facility under the same contract; and by the U.S. National Science Foundation, Division of Astronomical Sciences under Contract No. AST-0950945 to NOAO.

%%% 2MASS %%%

This publication makes use of data products from the Two Micron All Sky Survey, which is a joint project of the University of Massachusetts and the Infrared Processing and Analysis Center/California Institute of Technology, funded by the National Aeronautics and Space Administration and the National Science Foundation.

%%% WISE %%%

This publication makes use of data products from the Wide-field Infrared Survey Explorer, which is a joint project of the University of California, Los Angeles, and the Jet Propulsion Laboratory/California Institute of Technology, funded by the National Aeronautics and Space Administration.

This publication also makes use of data products from NEOWISE, which is a project of the Jet Propulsion Laboratory/California Institute of Technology, funded by the Planetary Science Division of the National Aeronautics and Space Administration.

%%% Chandra %%%

The scientific results reported in this article are based in part on data obtained from the Chandra Data Archive.

\end{acknowledgements}

\bibliography{main}{}
\bibliographystyle{aasjournalv7}

\end{document}